\documentclass[12pt]{article}
\usepackage{lineno} 
\usepackage{lipsum}
\usepackage{multirow}
\usepackage{mathpazo}
\usepackage[utf8]{inputenc}
\usepackage{xcolor, graphicx}
\usepackage{framed}
\usepackage{makecell}
\usepackage{tikz}
\usepackage{arydshln} 
\usepackage[flushleft]{threeparttable} 
\usepackage{booktabs}
\usepackage[font=small]{caption}
\usepackage{graphicx}
\usepackage{tabularx} 
\usepackage[ruled]{algorithm2e}
\usepackage{xr-hyper}

\definecolor{darkgreen}{rgb}{0.0, 0.26, 0.15}

\usepackage[colorlinks,citecolor=blue,urlcolor=blue,linkcolor=darkgreen]{hyperref}
\usepackage[citestyle=apa,style=apa,backend=biber,url=false,uniquename=false, useprefix=true, doi=false]{biblatex}
\usepackage{dsfont} 
\AtEveryBibitem{%
  \clearfield{note}%
}
\usepackage{dsfont} 
\AtEveryBibitem{%
  \clearfield{note}%
}
\makeatletter
\newcommand*{\addFileDependency}[1]{
\typeout{(#1)}
\@addtofilelist{#1}
\IfFileExists{#1}{}{\typeout{No file #1.}}
}\makeatother

\usepackage{mathtools, amsfonts, amssymb}
\usepackage{amsmath}
\numberwithin{equation}{section}
\mathtoolsset{showonlyrefs=true}
\allowdisplaybreaks[2]
\usepackage{enumerate}
\usepackage{siunitx}

\usepackage{color}
\usepackage[normalem]{ulem}

\providecommand{\keywords}[1]
{
  \small	
  \textit{Key words---} #1
}

\usepackage{numprint}
\usepackage{booktabs}

\usepackage{chngcntr}
\usepackage{apptools}
\AtAppendix{\counterwithin{lemma}{section}}

\newcommand{\boldbeta}{\boldsymbol{\beta}}
\newcommand{\boldut}{\mathbf{u}_i(t)}
\newcommand{\boldepst}{\boldsymbol{\varepsilon}_{ij}(t)}
\newcommand{\Expec}{\mathbb{E}}

\newcommand{\Var}{\operatorname{Var}}
\newcommand{\SE}{\operatorname{SE}}
\newcommand{\Cov}{\operatorname{Cov}}
\newcommand{\diag}{\operatorname{diag}}
\newcommand{\ICC}{\operatorname{ICC}}
\DeclareMathOperator{\vect}{vec}

\newcommand*\samethanks[1][\value{footnote}]{\footnotemark[#1]}

\newcommand{\pkg}[1]{{\normalfont\fontseries{b}\selectfont #1}} \let\proglang=\textsf \let\code=\texttt
\title{Analysing kinematic data from recreational runners using functional data analysis}
\author{Edward Gunning
\thanks{\textbf{Corresponding author: \href{mailto:edward.gunning@pennmedicine.upenn.edu}{edward.gunning@pennmedicine.upenn.edu}}}
\thanks{Department of Biostatistics, Epidemiology and Informatics, University of Pennsylvania}
\and
Steven Golovkine\thanks{MACSI, Department of Mathematics and Statistics, University of Limerick, Ireland}
\and
Andrew J. Simpkin\thanks{School of Mathematical and Statistical Sciences, University of Galway, Ireland}
\and
Aoife Burke\samethanks[6]
\and
Sarah Dillon\samethanks[6] \samethanks[7] \thanks{School of Allied Health, Faculty of Education and Health Science, University of Limerick, Limerick, Ireland}
\and
Shane Gore\samethanks[6] \samethanks[7]
\and
Kieran Moran\thanks{Centre for Injury Prevention and Performance, Athletic Therapy and Training; School of Health and Human Performance, Dublin City University, Dublin, Ireland}
\thanks{Insight SFI Research Centre for Data Analytics, Dublin City University, Dublin, Ireland}
\and
Siobhan O'Connor\samethanks[6]
\and
Enda Whyte\samethanks[6]
\and
Norma Bargary\samethanks[3]
}
\date{}

\begin{document}

\maketitle

\begin{abstract}
We present a multivariate functional mixed effects model for kinematic data from a large number of recreational runners ($N=288$).
The runners' sagittal plane hip and knee angles are modelled jointly as a bivariate function with random effects functions used to account for the dependence among measurements from either side of the body. 
The model is fitted by first applying multivariate functional principal component analysis (mv-FPCA) and then modelling the mv-FPCA scores using scalar linear mixed effects models. 
Simulation and bootstrap approaches are introduced to construct simultaneous confidence bands for the fixed effects functions, and covariance functions are reconstructed to summarise the variability structure in the data and thoroughly investigate the suitability of the proposed model.
In our scientific application, we observe a statistically significant effect of running speed on both the hip and knee angles.
We also observe strong within-subject correlations, reflecting the highly idiosyncratic nature of running technique.
Our approach is more generally applicable to modelling multiple streams of smooth kinematic or kinetic data measured repeatedly for multiple subjects in complex experimental designs.

\end{abstract}


\section{Introduction}

Advances in data collection, processing and storage technologies have led to an increased volume of data produced for biomechanics and human movement research \parencite{ferber_gait_2016}. Forces (kinetics) or displacement (kinematics) are measured hundreds or thousands of times per second during a single movement, leading to datasets characterised by high-dimensional observations.
Functional data analysis (FDA) \parencite{ramsay_functional_2005} is particularly well-suited to modelling human movement data as it treats a time series of kinetic or kinematic data as a single function (or curve) rather than as a sequence of discrete measurements. This allows a more comprehensive analysis than reducing the time series to a single summary value (e.g., peak angle) or ignoring the time dependence in the high-dimensional sequences of measurements \parencite{hebert-losier_one-leg_2015, pataky_zero-_2015, warmenhoven_pca_2021}. 
Applications of FDA in biomechanics and human movement research include: describing the effects of orthoses on running or walking \parencite{coffey_common_2011, zhang_testing_2017}, clustering runners according to footfall pattern \parencite{liebl_ankle_2014}, predicting fatigue in recreational athletes \parencite{wu_predicting_2019} and classifying different forms of activity \parencite{aguilera-morillo_multi-class_2020}.

Our motivating dataset comes from the Dublin City University (DCU) Running Injury Surveillance Centre (RISC) study, which aims to investigate the relationship between clinical and biomechanical variables and running-related injuries (RRIs) among novice and recreational runners.
{Although recreational running is one of the most popular recreational hobbies in the world and it provides substantial positive benefits for health and well-being, RRIs present a considerable barrier to participation and several other negative consequences, e.g., negative health aspects and financial costs \parencite{hespanhol_junior_health_2017}.
Despite this, our understanding of RRIs is limited, especially with respect to biomechanical factors, which has motivated studies to investigate the relationship between biomechanical variables and RRIs in populations of recreational runners.
In particular, there has been a large focus on the population of recently-injured runners \parencite[see, e.g.,][]{bramah_is_2018, becker_biomechanical_2017, mann_association_2015}, as history of a recent RRI is the strongest risk factor for suffering a new one.
It is hypothesised that recently-injured runners might retain some of the movement characteristics that contributed to the previous injury, or adopt compensatory mechanisms that cause them to be re-injured \parencite{saragiotto_what_2014}.}
Findings of these studies have largely been conflicting, in part perhaps because they have employed traditional statistical techniques using discrete kinematic variables \parencite[e.g.,][]{ceyssens_biomechanical_2019, willwacher_running-related_2022}.
{The ability of FDA methods to preserve the salient structure in time-dependent biomechanical data could lead to more comprehensive analyses that improve our understanding of RRIs and biomechanical factors.}

\begin{figure}[h]
    \centering
    \includegraphics[page=1, width = 1\textwidth]{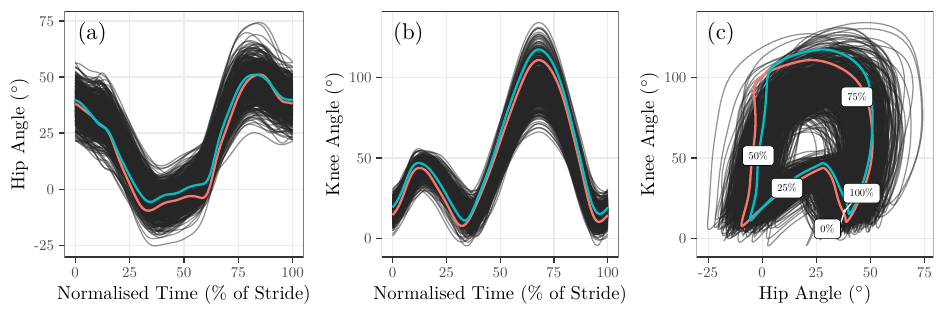}
    \caption{The dataset used in this analysis. \textbf{(a)} The hip angle functions. \textbf{(b)} The knee angle functions. \textbf{(c)} The knee angle functions plotted against the hip angle functions in an angle-angle diagram. In each plot, the right and left side observations for a single participant are highlighted in turquoise and red, respectively. The data have been time normalised and registered in preparation for analysis as described in Section \ref{data-prep}, {and are evaluated on a grid of $101$ points $t = 0, 1, \dots, 100$ for plotting}.}
    \label{fig:intro-plot}
\end{figure}

{Male and female runners between $18$ and $64$ years of age participated in the RISC study.
Whole-body kinematic data were recorded during a three-minute treadmill run, where the participant ran at a self-selected speed that reflected their typical training pace.
In addition, they completed a survey detailing their demographics, injury history (i.e., retrospective injury information) and training habits and were monitored for the occurrence of RRIs for a 12-month period (i.e., prospective injury information); see Table \ref{tab:tab1chpt3.} for summary characteristics of the participants in the dataset.
For this dataset, the relationship between injury history and scalar clinical \parencite{dillon_injury-resistant_2021} and scalar biomechanical \parencite{burke_comparison_2022} variables has been examined, but approaches that preserve the full biomechanical time series data have not been employed.
Focusing on the hip and knee angles in the sagittal plane (Figure \ref{fig:intro-plot}), we aim to to characterise the effect of retrospective injury status on the full biomechanical time series, while accounting for and understanding the effects of other factors, e.g., sex, running speed and age.
\emph{Function-on-scalar regression models} \parencite{faraway_regression_1997, ramsay_functional_2005} are an appropriate tool for characterising these relationships, where the biomechanical time series' are treated as the functional response variable(s), modelling their dependence on scalar covariates, e.g., injury status, sex, running speed and age. \textcite[Section 5]{morris_functional_2015} provides a comprehensive review of conventional function-on-scalar regression models.}

{Conventional function-on-scalar regression models assume independent observations, and do not handle dependence induced by repeated observations from the same individual.
However, these dependencies frequently arise in biomechanics for a number of reasons, e.g., multiple strides, trials or repetitions of a movement, or measurements from both sides of the body. 
In our case, although we have computed an average of all strides on the right and left side separately (Figure \ref{fig:intro-plot}), further averaging across the right and left sides to produce a single bilateral average curve could lead to a substantial loss of information and it could potentially bias subsequent analyses if large asymmetries exist.
\emph{Functional mixed effects} (or \emph{multilevel}) models, which are the analogue of classical scalar mixed effects models \parencite{laird_random-effects_1982, bates_fitting_2015}, extend conventional function-on-scalar regression models to handle repeated measures settings and more complex dependence structures.
The literature on functional mixed effects models is rich -- early pioneering work was by \textcite{morris_wavelet-based_2006, guo_functional_2002}, later developments by \textcite{scheipl_functional_2015, cui_fast_2022}, reviews are provided by \textcites[Section 5.7]{morris_functional_2015}{liu_functional_2012}{morris_comparison_2017} and a recent application in running biomechanics by \textcite{matabuena_estimating_2023}.}

\begin{table}
\centering
\begin{tabular}[t]{llrr}
\toprule
  &    & \textbf{Mean} & \textbf{Std. Dev.}\\
\midrule
Speed (\si{\km \per \hour}) &  & 11.0 & 1.6\\
Age (years) &  & 43.3 & 9.0\\
Weight (kg) &  & 72.4 & 12.9\\
Height (cm) &  & 172.9 & 9.7\\
\midrule
 &  & \textbf{N} & $\mathbf{\mathbf{(\%)}}$\\
\midrule
Retrospective Injury Status & Never Injured & 50 & 17.4\\
 & Injured $>2$ yr. ago & 67 & 23.3\\
 & Injured $1-2$ yr. ago & 51 & 17.7\\
 & Injured $<1$ yr. ago & 120 & 41.7\\
Sex & Male & 176 & 61.1\\
 & Female & 112 & 38.9\\
\bottomrule
\end{tabular}
\caption{Summary characteristics of the participants in the RISC dataset included in this analysis.}
\label{tab:tab1chpt3.}
\end{table}

{Rather than fitting separate (univariate) functional mixed effects models to the data from the knee and hip, it makes sense from a methodological and applied perspective to model them collectively.
From a statistical perspective, sharing information among functional variables can lead to improved parameter estimates \parencite{volkmann_multivariate_2021, zhu_multivariate_2017}, and from a biomechanical perspective it is preferable to model and interpret the knee and hip jointly (Figure \ref{fig:intro-plot} (c)) because they work together as parts of a system and understanding their interaction (i.e.,  coordination) is crucial \parencite{glazier_beyond_2021}.
\emph{Multivariate} functional data analysis techniques \parencite[see, e.g.,][]{gorecki_selected_2018} concern the analysis of multiple functional variables (e.g., the knee and hip angles), and they have been shown to be useful for understanding co-ordination among multiple joints in sports biomechanics \parencite{ryan_functional_2006,trounson_effects_2020}.
\emph{Multivariate} (or \emph{multiple-response}) functional mixed effects models extend classical univariate functional mixed effects models to handle multiple functional variables as outcomes.
However, the literature on these models is more scarce than in the univariate case\footnote{Methods for multivariate functional regression and inference \emph{without} random effects/ multilevel structures have been developed by \textcite{jiang_analysis_2022, diquigiovanni_conformal_2022, zhu_one-way_2022, liu_multivariate_2022, li_latent_2023}.}, with just three main approaches proposed \parencite{goldsmith_assessing_2016, volkmann_multifamm_2021, zhu_multivariate_2017} and they have yet to be applied in human-movement/running biomechanics.}

{\textcite{goldsmith_assessing_2016} developed a bespoke bivariate functional mixed effects model for kinematic data from a motor control experiment, where linear fixed effects of scalar covariates and subject-specific random effects were modelled using penalised splines. Their model was fitted in a Bayesian framework (using both variational approximations and full Markov Chain Monte Carlo (MCMC) sampling).
\textcite{volkmann_multivariate_2021} proposed an alternative approach, by extending the univariate Functional Additive Mixed Model (FAMM) to the multivariate setting. 
In this model, smooth non-linear effects of scalar covariates and multiple layers of random effects were modelled using penalised splines and multivariate Functional Principal Components (mv-FPCs), respectively.
It is fitted in a Frequentist framework by recasting the functional model as a large scalar additive mixed model and using the \pkg{mgcv} software \parencite{wood_fast_2011}, readily accommodating functions that are sparsely or irregularly measured with error.
Finally, \textcite{zhu_multivariate_2017} extended the Bayesian Functional Mixed Model (BayesFMM) \emph{basis modelling} approach of \textcite{morris_wavelet-based_2006} to handle multivariate functional data.
Their approach involves projecting each multivariate functional observation onto a set of basis functions and then modelling each basis coefficient separately using Bayesian scalar linear mixed effects models.
This ``divide and conquer" strategy makes it scalable to large datasets and facilitates the specification of a variety of complex random effects structures.
Therefore, we use the general approach of \textcite{zhu_multivariate_2017} to model the RISC dataset, with modifications that are motivated by the application at hand.}

{In particular, we present a Frequentist implementation of the Bayesian basis modelling approach, which was noted as a possible extension by \textcite{zhu_multivariate_2017} but not pursued.
This allows the model to be fitted using existing open-source mixed effects modelling software.
However, it does not produce posterior samples for pointwise and simultaneous inference of fixed effects, so for this we adapt existing Frequentist resampling and simulation techniques.
The basis modelling approach makes the assumption that each basis coefficient can be modelled separately, though the suitability of this assumption is not always checked in practice.
As such, we present an approach to graphically assess the suitability of this assumption for our application by comparing covariance reconstructions to unstructured estimates.
Finally, we extend the intraclass correlation coefficient (ICC) for univariate functional data \parencite{di_multilevel_2009} to the multivariate case, to summarise the degree of intra-subject correlation in our application.}

The remainder of the article is structured as follows.
In Section \ref{sec:bfmm-model}, we describe our proposed methodology and its implementation.
Section \ref{sec:bfmm-results} contains the data analysis and results of our scientific application. We close with a discussion in Section \ref{sec:bfmm-discussion}.
{A simulation study, additional methodological and application details, and a sensitivity analysis using alternative modelling approaches are contained in a supplementary appendix.}

\section{Methodology}\label{sec:bfmm-model}
\subsection{Model}

We denote the bivariate functional observation from the $i$th individual on side $j$ as
$$
\mathbf{y}_{ij} (t) = (y_{ij}^{(hip)} (t), y_{ij}^{(knee)} (t))^\top,
$$
for $i = 1, \dots, N$ where $N$ is the total number of individuals, $j \in \{\text{left, right}\}$ and $t \in [0, T]$ which is a normalised time interval. 
We let $\mathbf{x}_{ij} = (x_{ij1}, \dots, x_{ijA})^\top$ denote the vector of length $A$ of scalar covariates for  subject $i$ on side $j$. The covariates could be subject specific (e.g., sex, height) or subject-and-side specific (e.g., an indicator for a subject's dominant side).

The bivariate functional mixed effects model is
\begin{equation} \label{FMM}
    \mathbf{y}_{ij} (t) = \boldbeta_0 (t) + \sum_{a=1}^A x_{ija}  \boldbeta_a (t) + \boldut + \boldepst,
\end{equation}
where the bivariate function $\boldbeta_0 (t)~=~(\beta_0^{(hip)}(t), \beta_0^{(knee)}(t))^\top$ is the intercept function, the bivariate function $\boldbeta_a (t)~=~(\beta_a^{(hip)} (t), \beta_a^{(knee)} (t))^\top$ is the fixed effect regression coefficient function corresponding to the $a$th covariate, the bivariate function $\boldut~=~(u_i^{(hip)} (t), u_i^{(knee)} (t))^\top$ is the functional random intercept for the $i$th subject and $\boldepst~=~(\varepsilon_{ij}^{(hip)} (t), \varepsilon_{ij}^{(knee)} (t))^\top$ is the functional random error specific to the $i$th subject on side $j$.

The model is the bivariate functional analogue of a scalar linear mixed effects model with a single grouping variable \parencites[e.g.,][]{laird_random-effects_1982}. 
For $a~=~1, \dots, A$, the fixed-effect function $\boldbeta_a (t)$ captures how the $a$th scalar covariate influences the ``expected level and shape" of the bivariate functional response \parencite{bauer_introduction_2018}.
We assume that the bivariate functional random intercepts $\boldut,~i~=~1,\dots,N$ are independent realisations of a zero-mean multivariate Gaussian process with a matrix-valued covariance function $\mathbf{Q}$. The bivariate functional random intercepts take into account the grouping structure in the data, i.e., that the left and right side hip and knee angle functions from the same subject are likely to be similar and should share a subject-specific average function.
{A standard multivariate function-on-scalar regression model without these random effects would ignore this intra-subject correlation, effectively treating an individual's observations from the right and left side as independent.}
Analogous to random intercepts in scalar linear mixed models, they can be thought of as capturing the correlation between observations from the same subject, or accounting for average differences between subjects. 
We assume that the bivariate functional random errors $\boldepst,~i~=~1,\dots,N$, $j \in \{\text{left, right}\}$ are independent realisations of a zero-mean multivariate Gaussian process with a matrix-valued covariance function $\mathbf{S}$. They are often referred to as ``curve-level functional random effects'' because they capture correlation within, rather than between, functional observations \parencite{morris_functional_2015}.

We stack all functional terms in the model to give
\begin{equation} 
    \mathbf{Y} (t) = \mathbf{XB} (t) + \mathbf{ZU}(t) + \mathbf{E} (t), \label{Matrix-FMM}
\end{equation}
where the matrix $\mathbf{Y}(t)~=~(\mathbf{y}_{1 \text{,left}} (t) \ | \ \cdots \ | \ \mathbf{y}_{N \text{,right}} (t))^\top$ represents the functional observations, the matrix $\mathbf{B}(t)~=~( \boldbeta_0 (t) \ | \ \cdots \ | \ \boldbeta_A (t))^\top$ represents the functional fixed effects, the matrix $\mathbf{U}(t)~=~(\mathbf{u}_1(t)\ | \ \cdots \ | \ \mathbf{u}_N(t))^\top$ represents the functional random effects, the matrix $\mathbf{E}(t)~=~(\boldsymbol{\varepsilon}_{1 \text{,left}}(t) \ | \ \cdots \ | \ \boldsymbol{\varepsilon}_{N \text{,right}} (t))^\top$ represents the functional random errors and $\mathbf{X}$ and $\mathbf{Z}$ are $2N \times (A+1)$ and $2N \times N$ design matrices for the fixed and random effects, respectively. Written in this way, the model is a bivariate version of the general functional mixed model \parencite{morris_wavelet-based_2006}.

\begin{figure}
    \centering
    \includegraphics[width=1\linewidth]{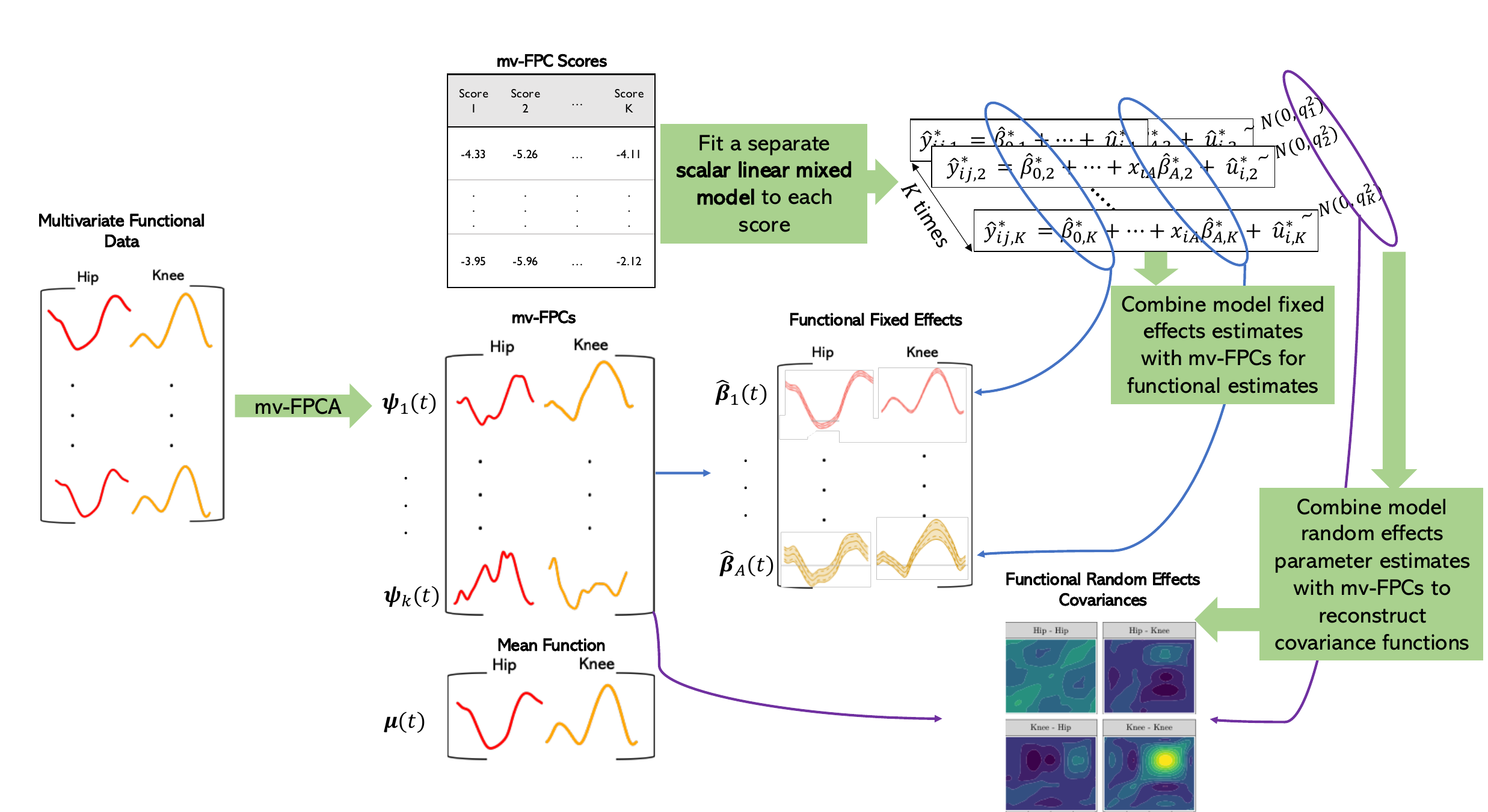}
    \caption{A flowchart summarising the main sequence of steps in our modelling approach.}
    \label{fig:flowchart}
\end{figure}

Our approach for fitting the model, described in the remainder of this section, can be summarised as follows. 
{First, the multivariate functional data are expanded on a basis of multivariate functional principal components (mv-FPCs) (Section \ref{sec:basis-expansion}).}
Scalar mixed models are fitted to each of the resulting FPC scores independently (Section \ref{sec:bfmm-estimation}). Estimates of the model parameters are combined across the bivariate FPCs to give estimates of the functional model terms (Sections \ref{sec:bfmm-fixef} and \ref{sec:bfmm-ranef}).
{The main steps are also summarised graphically in Figure \ref{fig:flowchart}.}
{A short simulation study to assess this approach in realistic data-generating scenarios is contained in Appendix \ref{sec:bfmm-simulation}}.

\subsection{Basis Expansion}\label{sec:basis-expansion}

{The first step in the basis modelling strategy is to represent each individual observation using a \emph{basis expansion}, that is
$$
\mathbf{y}_{ij} (t) = \sum_{k=1}^K y^*_{ijk} \boldsymbol{\psi}_k (t),
$$
where $\{y^*_{ijk}\}_{k=1}^K$ are scalar basis coefficients and $\{ \boldsymbol{\psi}_k (t) = (\psi_k^{hip}(t), \psi_k^{knee}(t))^\top \}_{k=1}^K$ are bivariate basis functions. As noted by \textcite{ramsay_functional_2005} and \textcite{morris_functional_2015}, the type of basis chosen should suit the characteristics of the data at hand. 
Popular choices for univariate functions are wavelets or FPCs \parencite{morris_wavelet-based_2006, aston_linguistic_2010, lee_bayesian_2019}.
For multivariate functional data, such as the kinematic data in our application, multivariate FPCs are a natural choice of basis, because they capture common variation across the dimensions of the multivariate functional data.
That is, in our application, mv-FPCA produces a set of basis functions that are useful for capturing variation in the hip and knee jointly.

Implicitly, this is the same type of basis that \textcite{zhu_multivariate_2017} used to represent the multivariate functional data in their work. 
Though they did not explicitly mention mv-FPCA, they expanded the functional data within each dimension on an orthonormal univariate basis and then performed a standard PCA of the combined matrix of basis coefficients, using the resulting PCA scores as new basis coefficients in the modelling.
As demonstrated by \textcite{happ_multivariate_2018}, this is one approach to estimating mv-FPCs.
However, there are other approaches to estimating the mv-FPCs that are equally valid.
In our application, we chose to construct the mv-FPC basis from univariate B-spline (non-orthonormal) basis expansions within each dimension \parencite{jacques_model-based_2014, happ_multivariate_2018}, as B-splines are known to be suited to representing smooth functions with local features \parencite{ramsay_functional_2005, morris_functional_2015}.
The mv-FPCA basis could also be estimated directly from discrete observations of the multivariate functional data \parencites[][]{li_fast_2020}[p. 167]{ramsay_functional_2005}.}

The only requirement is that the basis expansion is \emph{near-lossless}, which according to \textcite[p. 72]{morris_comparison_2017}, means it is ``sufficiently rich such that for all practical purposes it can recapitulate the observed functional data". In general, this property is controlled by $K$, the number of mv-FPCs retained.
Because we calculate the mv-FPCA from univariate expansions, the richness of the basis additionally depends on the number of univariate basis functions used within each dimension.
We use a large number of univariate basis functions within each dimension, as the estimated mv-FPCA has been shown to be sensitive to using too few univariate basis functions \parencite{golovkine_use_2023}.
When choosing $K$, the number of mv-FPCs to keep, retaining a larger number will give a closer fit to the observed data, while retaining fewer mv-FPCs (``truncation'' or ``compression'') makes the representation smoother and reduces computation time taken to model the scores.
As our data are smooth and we assume that they are measured without error, we choose a large number of mv-FPCs to explain a high variance-explained threshold of $99.99\%$ \parencite{zhu_multivariate_2017}.
{However, if we wanted to induce regularisation and avoid over-fitting, we could more carefully choose $K$ through cross-validation for near-lossless basis expansions \parencite[see, e.g.,][Chapter 2]{zohner_feature_2021}.}

We let $\mathbf{\Psi} (t)$ denote the $K \times 2$ matrix containing the mv-FPCs
$$
\mathbf{\Psi} (t) =
\begin{pmatrix}
\psi_1 ^{(hip)} (t)  & \psi_1 ^{(knee)} (t) \\
\vdots & \vdots\\
\psi_{K} ^{(hip)} (t)  & \psi_{K} ^{(knee)} (t) \\
\end{pmatrix},
$$
so that we can write $\mathbf{Y}(t) = \mathbf{Y}^* \mathbf{\Psi} (t)$, where $\textbf{Y}^*$ is the matrix of basis coefficients (i.e., mv-FPCA scores) which is obtained by projecting the $\textbf{Y}(t)$ onto the mv-FPCs
$$
\mathbf{Y}^* = \int_0^{T} \textbf{Y}(t) \boldsymbol{\Psi}(t)^\top \mathrm{d}t.
$$
The central idea of the basis modelling paradigm is to use the same basis for all terms in the model \eqref{Matrix-FMM}, i.e., $\mathbf{B} (t) = \mathbf{B}^* \mathbf{\Psi} (t)$, $\mathbf{U} (t) = \mathbf{U}^* \mathbf{\Psi} (t)$ and $\mathbf{E} (t) = \mathbf{E}^* \mathbf{\Psi} (t)$ so that the following ``basis-space'' model can be fitted instead
\begin{equation} \label{Matrix-ScalarMM}
    \mathbf{Y^*} = \mathbf{XB}^* + \mathbf{ZU}^* + \mathbf{E}^*,
\end{equation}
{which is obtained by projecting both sides of Equation \eqref{Matrix-FMM} onto $\boldsymbol{\Psi} (t)$ \parencite[see, e.g.,][Section 3.1.4]{morris_comparison_2017}.}
This simplifies the task from fitting a bivariate functional mixed model (the ``data-space'' model) to fitting a multivariate scalar linear mixed model (the ``basis-space'' model). 
Because mv-FPCA scores are (marginally) uncorrelated, we attain further simplification by assuming that the columns of $\mathbf{Y}^*$ are approximately independent and can be modelled separately.
This reduces the problem to fitting a series of univariate scalar linear mixed models to the columns of $\mathbf{Y}^*$, reducing computation times and memory requirements.

\subsection{Estimation} \label{sec:bfmm-estimation}
A {Gaussian} scalar linear mixed effects model is fitted separately to each FPC score, i.e., each column of $\mathbf{Y}^*$. The model for the $k$th basis coefficient, $k=1, \dots, K$, is 
\begin{equation}\label{eq:model_k}
y_{ijk}^*  = \beta_{0k}^*  + \sum_{a=1}^A x_{ija}  \beta_{ak}^* + u_{ik}^* + \varepsilon_{ijk}^{*},
\end{equation}
where $u^*_{ik} \stackrel{i.i.d.}{\sim} \mathcal{N}(0, q_k)$ and $\varepsilon_{ijk}^* \stackrel{i.i.d.}{\sim} \mathcal{N}(0, s_k)$.
{Here, the Gaussian specification for the random effects and random errors follows from the assumption of a Gaussian process for $\mathbf{u}_i(t)$ and $\boldsymbol{\varepsilon}_{ij}(t)$.}
The model can be estimated using either Bayesian or frequentist methods; {\textcite{zhu_multivariate_2017} took a Bayesian approach using custom a MCMC algorithm written in \proglang{MATLAB} and \proglang{C}}. We opt for a frequentist approach because it allows a fast and straightforward implementation using standard open-source software; we use the \code{lmer()} function from the \pkg{lme4} \parencite{bates_fitting_2015} \proglang{R} \parencite{r_core_team_r_2022} package to fit the models using REML. An introduction to REML estimation of linear mixed models is given by \textcite[Section 2.4.5]{wood_generalized_2017} and \textcite[Section 2.2.5]{pinheiro_mixed-effects_2006}. 


Implicitly, modelling each basis coefficient separately assumes that $\Cov(u^*_{ik}, u^*_{ik'}) \allowbreak = 0$ and $\Cov(\varepsilon_{ijk}^*, \varepsilon_{ijk'}^*) = 0$ for $k \neq k'$.
Although the mv-FPCA step produces basis coefficients that are marginally uncorrelated, the projections of the individual processes $\mathbf{u}_i(t)$ and $\boldsymbol{\varepsilon}_{ij}(t)$ onto the mv-FPCA basis are not guaranteed to be uncorrelated across $k$. 
However, this assumption is commonly made in basis modelling approaches for functional mixed models because it brings about simplifications in modelling and computation while maintaining a high degree of flexibility \parencite{aston_linguistic_2010, lee_bayesian_2019, zhu_multivariate_2017}. 
In Section \ref{sec:bfmm-covariance-reconstruction}, we describe a way to graphically assess the extent to which this assumption is reasonable for each process, based on the reconstruction of their respective covariance functions.


\subsection{Fixed Effects}\label{sec:bfmm-fixef}

We do not try to interpret the individual models fitted to the basis coefficients. Instead we combine the estimated parameters across coefficients with the basis functions to reconstruct the functional model terms. This step is referred to as ``transforming the estimates back to the data space''. For a given fixed-effect function $\boldbeta_a (t)$, we use the estimates $\widehat{\beta}_{ak}^*$ of $\beta_{ak}^*$, $k=1,\dots, K$ to construct an estimate
$$
\widehat{\boldbeta}_a (t)  = \sum_{k=1}^{K} \widehat{\beta}_{ak}^* \boldsymbol{\psi}_k (t) = \mathbf{\Psi} (t)^\top \widehat{\boldbeta}_a^*, \quad \text{where} \ \ \widehat{\boldbeta}_a^* = (\widehat{\beta}_{a1}^*, \dots, \widehat{\beta}_{aK}^*)^\top.
$$

\subsubsection{Pointwise Confidence Intervals}\label{sec:bfmm-pw-conf-int}

Pointwise confidence intervals for $\boldbeta_a (t)$ can be constructed based on a Gaussian approximation $\widehat{\boldbeta}_a ^* \sim \mathcal{N}_{K}({\boldbeta}_a, \widehat{\boldsymbol{\Sigma}}_a)$, where $\widehat{\boldsymbol{\Sigma}}_a = \diag\left\{\widehat{\Var}\left(\widehat{\beta}_{a1}^*\right), \dots, \widehat{\Var}\left(\widehat{\beta}_{aK}^*\right)\right\}$. This gives the pointwise variance function
$$
\widehat{\Var}(\widehat{\boldbeta}_a (t)) \approx \mathbf{\Psi} (t)^\top \widehat{\boldsymbol{\Sigma}}_a \mathbf{\Psi} (t),
$$
so that an approximate pointwise confidence interval can be constructed as $\widehat{\boldbeta}_a (t) \pm q_{1-\alpha/2} \times \widehat{\SE}(\widehat{\boldbeta}_a (t))$ where $\widehat{\SE}(\widehat{\boldbeta}_a(t))$ is the square-root of $\widehat{\Var}(\widehat{\boldbeta}_a (t))$ and $q_{1-\alpha/2}$ is the $(1-\alpha/2)$th quantile of the standard Gaussian distribution. These are Wald intervals because they are based on the Gaussian approximation for each $\widehat{\beta}_{ak}^*$ and are only asymptotically valid because the estimate
$\widehat{\Var}(\widehat{\beta}_{ak}^*)$ is used in place of the true $\Var(\widehat{\beta}_{ak}^*)$ \parencite{kenward_small_1997}. Despite this, Wald intervals are quick and straightforward to compute and are returned by default by standard mixed model software. For our application with a large number of study participants the approximation should be reasonable so they are a convenient tool. In Section \ref{sim-conf-bands}, we describe a more computationally intensive bootstrap technique for constructing simultaneous confidence bands which can also be used to construct pointwise intervals.



\subsubsection{Simultaneous Confidence Bands}\label{sim-conf-bands}

Pointwise confidence intervals for bivariate functional parameters only provide coverage within a given dimension $p \in \{\text{hip, knee}\}$ at a specific point $t \in [0, T]$. They will not, in general, provide nominal coverage for the entire function $\boldbeta_a (t)$ because of the multiple-testing problem \parencite{degras_simultaneous_2017}. We define a simultaneous confidence band as the band $[\boldsymbol{\beta}_{a, L} (t),\ \boldsymbol{\beta}_{a, U} (t)]$ providing simultaneous coverage
$$
{P}( \beta_a^{(p)} (t) \in [\beta_{a, L}^{(p)} (t),\ \beta_{a, U}^{(p)} (t)], \forall \ p \in \{\text{hip, knee}\} \text{ and } t \in [0, T]) \approx 1 - \alpha.
$$
The band can be thought of as providing an adjustment for multiple testing along the whole domain $[0, T]$ and across the hip and knee dimensions.

Resampling or simulation techniques are typically used to build simultaneous confidence bands. A sketch of the general procedure first introduced by \textcite[Section 6.5]{ruppert_semiparametric_2003} for scatterplot smoothing, which has been subsequently been shown to work well for univariate functional data \parencite{crainiceanu_bootstrap-based_2012, park_simple_2018, cui_fast_2022}, is given in Algorithm \ref{alg:SCB}. The algorithm admits a number of ways to construct the bands in our application, differing in how the samples $\widehat{\boldbeta}_{a(1)} (t), \dots \widehat{\boldbeta}_{a(R)} (t)$ and the estimate $\widehat{\SE}(\widehat{\boldbeta}_a (t))$ are obtained (Step 1). The Wald approximation in Section \ref{sec:bfmm-pw-conf-int} can be used to simulate samples from $\mathcal{N}_{K}(\widehat{{\boldbeta}}_a, \widehat{\boldsymbol{\Sigma}}_a)$. Alternatively, parametric or non-parametric bootstrap techniques can be used to obtain the samples and estimate $\widehat{\boldsymbol{\Sigma}}_a$. We opt for the non-parametric bootstrap, where bootstrap samples are created by resampling subject indices with replacement, hence called the ``bootstrap of subjects" \parencite{crainiceanu_bootstrap-based_2012, park_simple_2018, cui_fast_2022}. Each time a subject appears in a bootstrap sample, they are assigned a new pseudo-ID which is used in model estimation. We use the bootstrap to estimate $\mathbf{\Sigma}_a$ and then draw samples from $\mathcal{N}_{K}(\widehat{{\boldbeta}}_a,\widehat{\boldsymbol{\Sigma}}_a)$ for Step 1 of Algorithm \ref{alg:SCB}, however the bootstrap samples could also be used directly \parencite{crainiceanu_bootstrap-based_2012}.


\begin{algorithm}
\caption{Level $\alpha$ simultaneous confidence bands for $\boldbeta_a (t)$ \parencite{crainiceanu_bootstrap-based_2012}.}\label{alg:SCB}
\KwData{$\widehat{\boldbeta}_a (t)$, $\widehat{\SE}(\widehat{\boldbeta}_a (t))$.}
\KwResult{Simultaneous confidence bands of  $\{\boldbeta_a (t), p \in \{\text{hip, knee}\} \text{ and } t \in [0, T] \}$.}
1. Obtain samples $\widehat{\boldbeta}_{a(1)} (t), \dots \widehat{\boldbeta}_{a(R)} (t)$ by simulation or bootstrap;

\For{$r = 1, \dots, R$}{
 2. Calculate $z_r = \max_{t, p} \{  \lvert \widehat{\boldbeta}_a (t) - \widehat{\boldbeta}_{a(r)} (t) \rvert / \widehat{\SE} (\widehat{\boldbeta}_a (t)) \}$;
}

3. Compute $z_{(1-\alpha)}$, the $(1-\alpha)$th empirical quantile of
$\{z_1, \dots, z_R\}$;

4. The simultaneous confidence band is calculated as 
$$
\widehat{\boldbeta}_a (t) \pm z_{(1-\alpha)} \widehat{\SE} (\widehat{\boldbeta}_a (t)).
$$
\end{algorithm}

\subsection{Random Effects}\label{sec:bfmm-ranef}

\subsubsection{Covariance Reconstruction}\label{sec:bfmm-covariance-reconstruction}

In scalar linear mixed effects models, we are not concerned with estimating the random effects themselves; instead we try to estimate the parameters that describe the random effects' distributions, i.e., the variance and covariance parameters \parencite[p. 195]{faraway_extending_2016}. Analogously in bivariate functional mixed effects models, we are concerned with estimation of the auto- and cross-covariance functions describing the bivariate functional random effects.

The bivariate functional random intercepts are given by
$$
\boldut = \sum_{k = 1}^{K} u_{ik}^* \boldsymbol{\psi}_k (t) = \mathbf{\Psi} (t)^\top  \mathbf{u}_i^{*},
$$
and due to the independence assumption for the basis coefficients, we have
$$
\Cov\left( \mathbf{u}_i^* \right) = \mathbf{Q}^* = \diag \{q_1, \dots, q_{K} \},
$$
where $q_1, \dots, q_{K}$ are random-intercept variances from the scalar mixed models (Section \ref{sec:bfmm-estimation}). Therefore, the matrix-valued covariance function for the bivariate functional random intercepts is given by
$$
\mathbf{Q}(t, t') = \Cov(\mathbf{u}_i (t), \mathbf{u}_i (t')) = \mathbf{\Psi} (t) ^\top  \mathbf{Q}^* \mathbf{\Psi} (t'), \quad t, t' \in [0, T].
$$
Similarly the matrix-valued covariance function for the bivariate functional random error is
$$
\mathbf{S}(t, t') = \Cov(\boldsymbol{\varepsilon}_{ij} (t), \boldsymbol{\varepsilon}_{ij} (t')) = \mathbf{\Psi} (t) ^\top  \mathbf{S}^* \mathbf{\Psi} (t'), \quad t, t' \in [0, T],
$$
where $\mathbf{S}^* = \diag\{s_1, \dots, s_{K}\}$. In practice, we replace $q_k$ and $s_k$ by their estimates $\widehat{q}_k$ and $\widehat{s}_k$ to obtain the reconstructions $\widehat{\mathbf{Q}}$ of $\mathbf{Q}$ and $\widehat{\mathbf{S}}$ of $\mathbf{S}$.

As mentioned in Section \ref{sec:bfmm-estimation}, the independence assumption for the basis coefficients restricts $\mathbf{Q}^*$ and $\mathbf{S}^*$ to be diagonal, limiting the types of covariance structures that can be estimated. \textcite{lee_bayesian_2019} recommend checking this assumption graphically by plotting the reconstructed covariance functions. 
For functions on large and possibly high-dimensional grids (e.g., images), it has typically not been feasible to compute unrestricted covariance estimates to compare the model reconstructions with.
In this work, we obtain fully unstructured estimates of the covariance functions by extending the multilevel FPCA method of \textcite{di_multilevel_2009} to multivariate functional data. 
By comparing the model and unstructured estimates graphically, we can assess whether the diagonal assumptions for $\mathbf{Q}^*$ and $\mathbf{S}^*$ are reasonable.
Full details on the calculation of the unstructured estimates are provided in Appendix \ref{unstructured-cov-est}.

\subsubsection{Functional Intraclass Correlation Coefficient}

Random-intercept scalar mixed models allow a partitioning of variability into between-subjects and within-subjects elements through the intraclass correlation coefficient (ICC) \parencite[Section 8.1]{faraway_extending_2016}. \textcite[Section 2.2]{di_multilevel_2009} extended the ICC to univariate functional data by integrating each term over the functional domain. We further extend it to multivariate functional data by integrating over the functional domain and summing over the dimensions. The multivariate functional $\ICC$ for our model is
$$
\ICC = \frac{\sum_{k=1}^{K} q_k}{
\sum_{k=1}^{K} q_k + \sum_{k=1}^{K} s_k}.
$$
In our application, it can be interpreted as the proportion of variability in the hip and knee angles (after accounting for fixed effects) attributable to differences between subjects. The remainder $(1-\ICC)$ represents the proportion attributable to differences within subjects between the left and right sides (asymmetry). Further details on the ICC are provided in Appendix \ref{app:icc}.

\section{Data Analysis and Results} \label{sec:bfmm-results}

\subsection{Data Preparation} \label{data-prep}

\subsubsection{Extraction, Segmentation and Alignment}\label{sec:bfmm-processing}

This section summarises the data collection, extraction and preparation for analysis. As per the Vicon Plug in Gait model (Vicon Motion Systems, Oxford, UK), 28 reflective markers (\SI{14}{\mm} in diameter) were placed at bony landmarks on the lower limbs, pelvis and trunk with an additional two markers placed on the anterior aspect of the mid tibia and mid thigh bilaterally. After a dynamic warm-up including treadmill running (FlowFitness, Runner-DTM2500i, Netherlands) for 6 minutes at a speed of \SI{9}{\km \per \hour}, participants completed a three-minute run at a self-selected pace that best represented their typical training pace. During the first minute of this three-minute run, kinematic data were collected using a 17-camera, three-dimensional motion analysis system (Vantage, Vicon, Oxford, UK) recording at 200\si{\hertz}. The marker trajectories were then filtered using a fourth-order zero-lag Butterworth filter at  15\si{\hertz}, chosen by residual analysis \parencite{winter_biomechanics_1979}. Functional joints and minimisation of soft tissue were calculated using the ``OSSCA" method in NEXUS 2 \parencite{taylor_repeatability_2010}. Sagittal plane hip and knee angles were then extracted bilaterally.

The extracted data were segmented into individual strides at the initial contact of the foot with the ground, which was identified as the first occurrence of two events: 1) the first negative vertical acceleration of the toe maker, and 2) the peak vertical acceleration of the heel marker. Both events were identified within a search window defined between the local maxima of the toe marker anterior position and the subsequent local minima of the ankle marker vertical position. For each stride, the time-argument values were then linearly re-scaled so that all curves shared the normalised domain $[0, 100]$, where $0$ represents the start of a stride and $100(\%)$ represents the end \parencite[i.e., linear time/ length normalisation,][]{helwig_methods_2011}.
{When discretisation of the functions was required, e.g., for plotting or computing the simultaneous bands, a grid of $101$ points $t = 0, 1, \dots, 100$ was used.}
Landmark registration \parencite{kneip_statistical_1992} was performed to further reduce timing variation in the functions. A single landmark was chosen to align the functional data from each stride -- the peak of the knee flexion angle. This landmark was chosen because it is clear and well-defined for every stride and easy to identify using a simple grid search. The hip and knee angles were registered simultaneously to this landmark to preserve the temporal correlation between them.

\subsubsection{mv-FPCA Calculation}

As described in Section \ref{sec:bfmm-processing}, the raw marker trajectories were filtered to remove observational error. Therefore, no additional smoothing was performed to avoid over-smoothing and dampening features in the data. Instead, the first-stage basis-function expansion interpolated, rather than smoothed, the data and reduced its dimension (i.e., reduced a large number of observation points, differing between curves, to a smaller number of common basis coefficients).

First, a B-spline basis was chosen to represent the univariate functional data in each dimension because it is a flexible basis and is well suited to smooth functions, such as the kinematic data at hand \parencite{ramsay_functional_2005, morris_functional_2015}. We found that $K_{hip} = K_{knee} = 80$ B-spline basis functions were sufficient to approximate the functional data from each stride almost perfectly. The basis coefficients were computed by ordinary least squares because no smoothing was required. Given the basis representation of the functional data for each individual stride, the reduced dataset of left and right side averages used in the analysis (Figure \ref{fig:intro-plot}) was obtained by averaging the basis coefficients of all strides for a given subject on a given side of the body.
{Computed from the univariate B-spline expansions}, the bivariate FPCA yielded $\widetilde{K}=38$ bivariate FPCs, satisfying the $99.99\%$ variance explained threshold. As expected, the majority of the variance was explained by the leading FPCs, e.g., $95\%$ of the variance was explained by the first seven FPCs, and $99\%$ by the first $13$. Additional information on the basis transformation is provided in Appendix \ref{additional-bfmm-basis-transformation}.

\subsection{Fixed Effects}
\begin{figure}
    \centering
    \includegraphics[width = 0.75\textwidth,page=10]{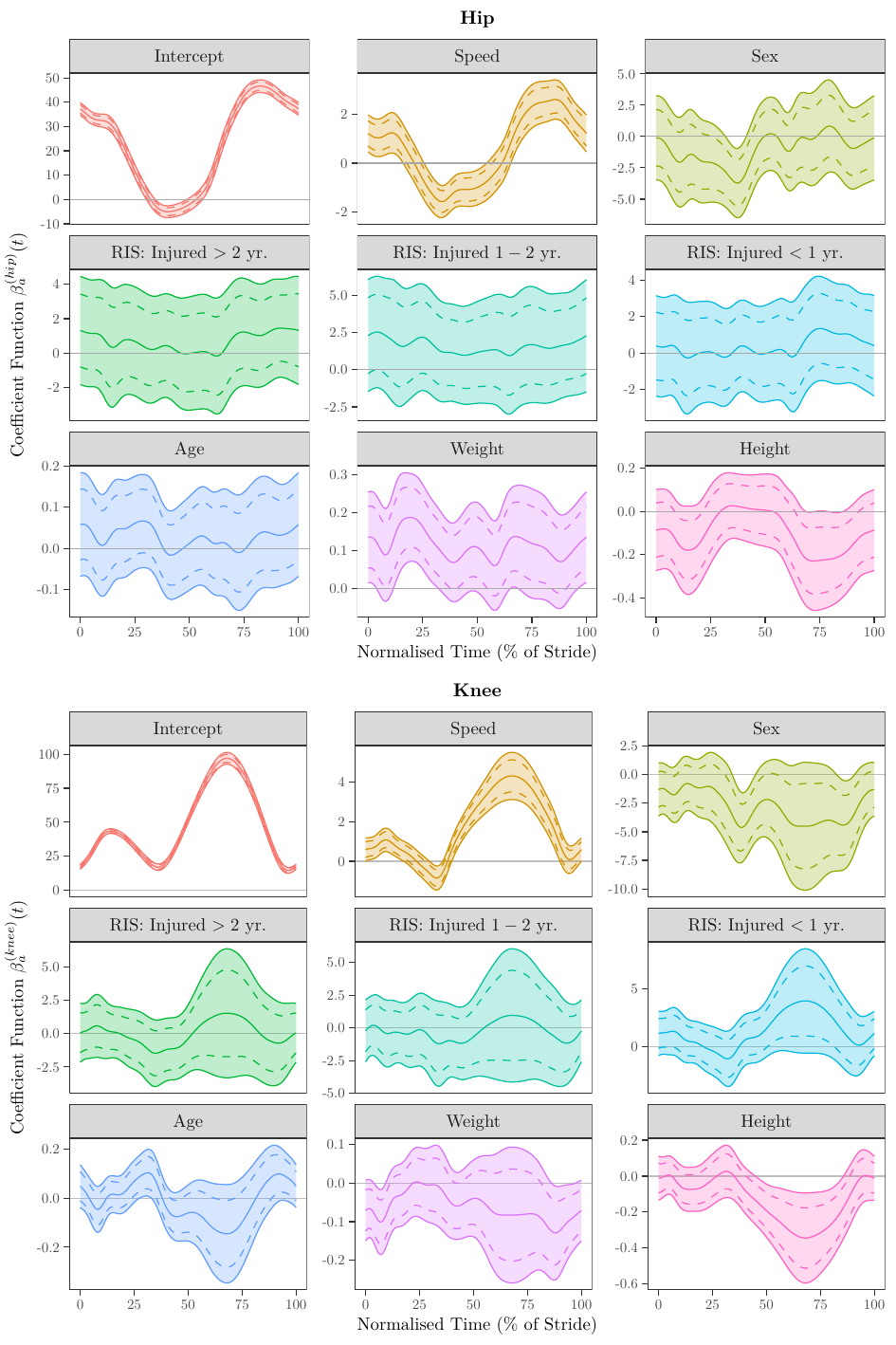}
    \caption{The estimated fixed-effect hip (top panel) and knee (bottom panel) regression coefficient functions. The solid line represents the point estimate function. The shaded ribbons represent $95\%$ simultaneous confidence bands obtained via bootstrap. The dashed lines represent $95\%$ pointwise confidence intervals obtained via bootstrap. Note: Wald confidence intervals are omitted because they are almost identical to those obtained with the bootstrap approach.}
    \label{fig:coef-funs}
\end{figure}

The fitted model was 
\begin{align}
    \mathbf{y}_{ij} (t) =& \ \boldbeta_0 (t) + \underbrace{\sum_{a=1}^3 x_{ia}  \boldbeta_a (t)}_{\text{Injury Status}} + \  \text{speed}_i \times \boldbeta_4 (t) + \text{sex}_i \times \boldbeta_5 (t) +  \text{age}_i \times   \boldbeta_6 (t) \\
    & + \text{weight}_i \times \boldbeta_7 (t) + \text{height}_i \times \boldbeta_8 (t) + \boldut + \boldepst,
\end{align}
where $x_{i1}, x_{i2}$ and $x_{i3}$ are dummy-coded variables representing the ``Injured more than 2 years ago", ``Injured 1-2 years ago" and ``Injured less than 1 year ago" categories of the retrospective injury status variable and the reference category is ``Never injured", $\text{speed}_i$ is the self-selected running speed of subject $i$ in \si{\km \per \hour}, $\text{sex}_i$ is a dummy-coded variable for sex of subject $i$ ($0=$ male, $1=$ female), $\text{age}_i$ is the age of subject $i$ in years, $\text{weight}_i$ is the weight of subject $i$ in kilograms and $\text{height}_{i}$ is the height of subject $i$ in centimetres. All numeric variables were centred to make the intercept function more interpretable. The regression coefficient functions for numerical and dummy-coded variables can be interpreted analogously to multiple linear regression; e.g., $\beta_4^{(hip)}(t)$ represents the expected change in the hip angle at $t$ for a 1-\si{\km \per \hour} increase in speed with all other variables held constant, and $\beta_5^{(knee)}(t)$ represents the expected difference in the knee angle at $t$ between females and males with all other variables held constant.

Figure \ref{fig:coef-funs} shows the estimated regression coefficient functions. The solid lines represent the point estimates, the shaded ribbons represent the $95\%$ simultaneous bands and the dashed lines represent $95\%$ pointwise confidence intervals. Results obtained via the Wald and bootstrap approaches were practically indistinguishable so only the bootstrap intervals are shown. The simultaneous bands are about 1.5 times as wide as the pointwise intervals. The confidence bands for the retrospective injury status regression coefficient functions contain zero (solid grey horizontal line) for all $t$, meaning that there is no evidence of a difference between any of the categories and the reference category of ``Never injured". Similarly, there is limited evidence of an age, height, weight or sex effect; although the simultaneous bands do not contain zero at certain points, the magnitude of each effect is small. However, self-selected speed has a strong effect in both the hip and knee dimensions -- the coefficient function has a distinct shape and the confidence band only contains zero when the function is changing from positive to negative. We turn to a more intuitive visualisation, based on model predictions, to interpret this effect.

\begin{figure}
    \centering
    \includegraphics[width = 0.85\textwidth]{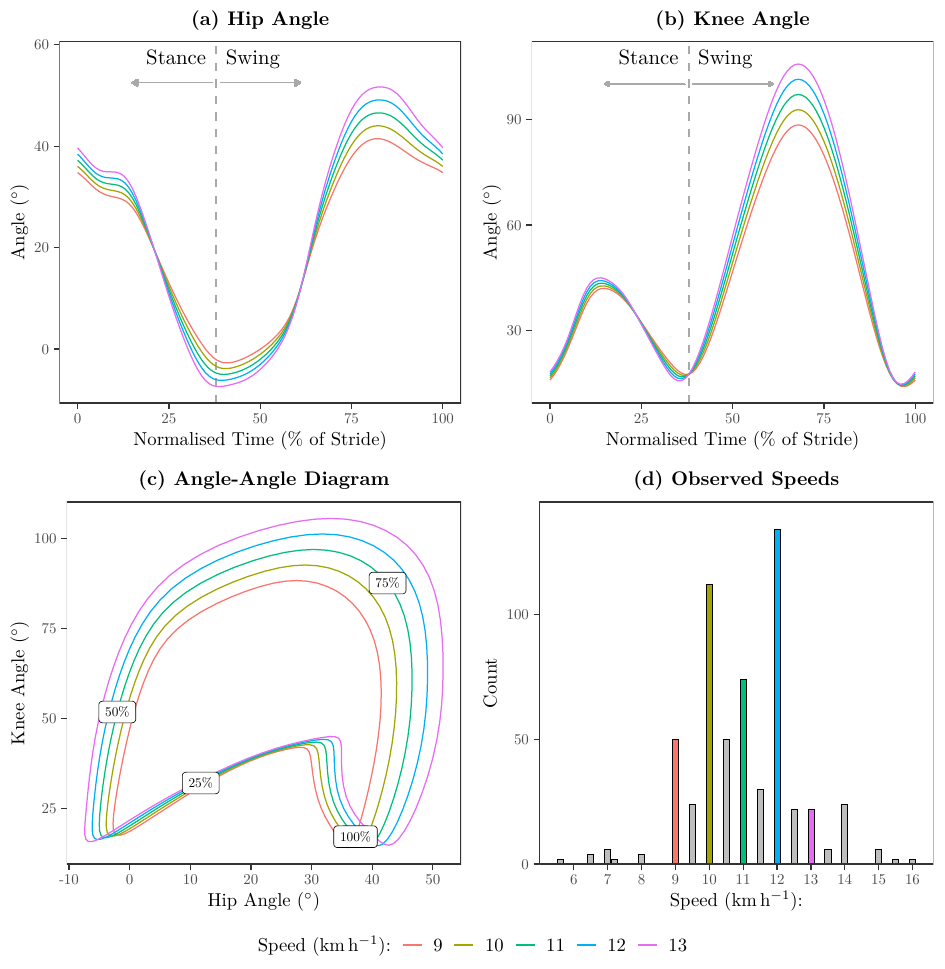}
    \caption{\textbf{(a)} Predicted values of the hip angle at different speeds plotted against time. \textbf{(b)} Predicted values of the knee angle at different speeds plotted against time. \textbf{(c)} Predicted values of the hip angle and knee angle at different speeds plotted against one another on an angle-angle diagram, with the approximate timings in the stride indicated with labels. \textbf{(d)} A barplot of the self-selected speed variable, with the speeds that were chosen for the visualisations in \textbf{(a)} -- \textbf{(c)} filled in their respective colours.
    \textbf{Note}: In plots \textbf{(a)} -- \textbf{(c)}, other numeric covariates are held at their mean value and categorical variables are set at their baseline/reference category.}
    \label{fig:speed-predictions}
\end{figure}

The regression coefficient functions, although useful for comparing effects and checking where confidence bands contain zero, do not give a representation of the estimated effect on the same scale as the observed data. Therefore, we predict the hip and knee angles at different speeds (while holding the other numeric variables at their mean and the categorical variables at baseline) and visualise the effects separately for each joint (Figure \ref{fig:speed-predictions} \textbf{(a)} and \textbf{(b)}) and in combination on an angle-angle diagram (Figure \ref{fig:speed-predictions} \textbf{(c)}). The observed effects are most evident in the swing phase of the movement ($t\approx38\%$ onward), where, on average, greater peak hip and knee flexion is associated with higher speeds.

\subsection{Random Effects}
\begin{figure}
    \centering
    \includegraphics[width = 1\textwidth,page=6]{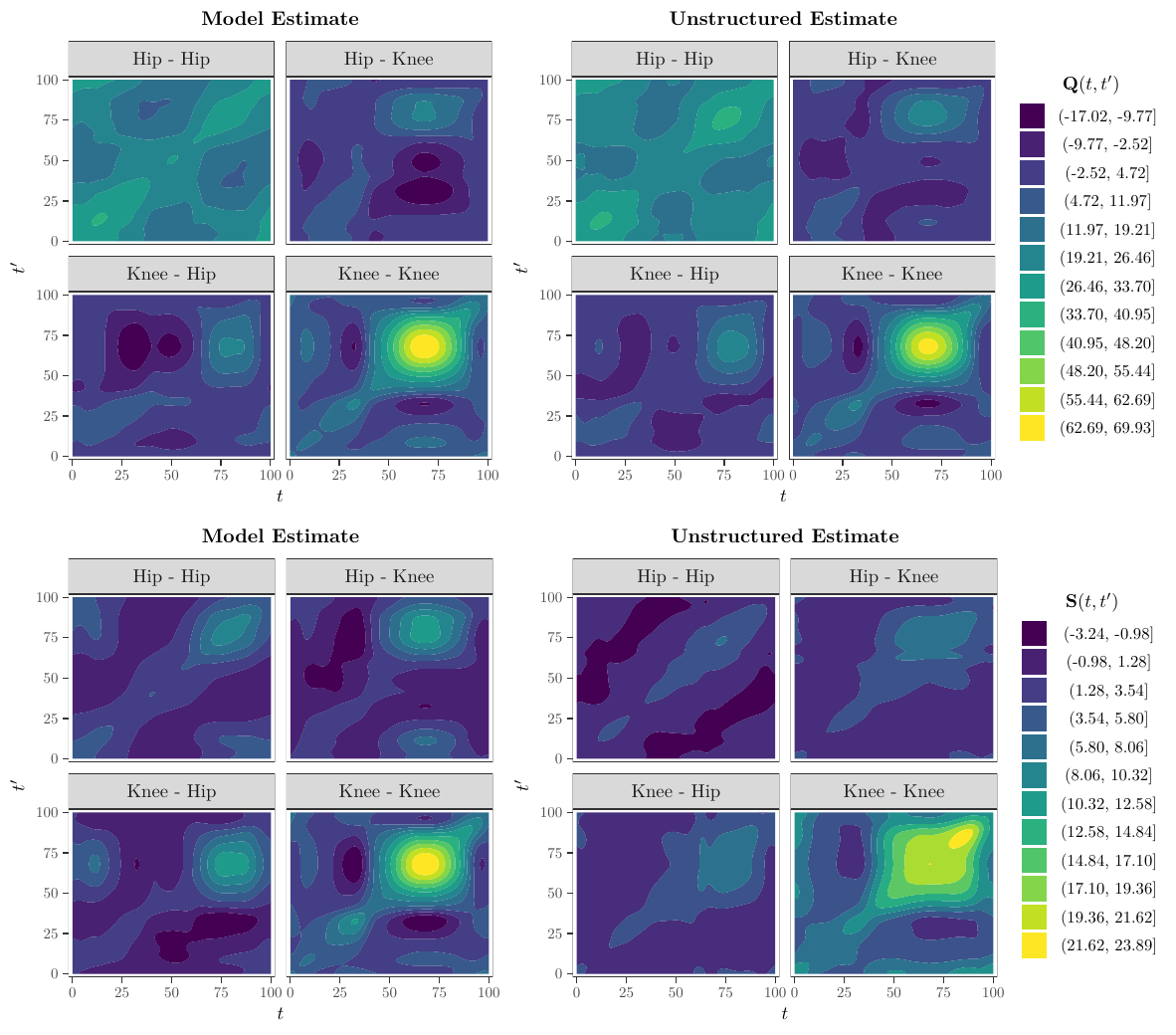}
    \caption{Filled-contour plots of the multivariate covariance functions. \textbf{Top panel}: The model (left) and unstructured (right) estimates of the multivariate functional random effects covariance function $\textbf{Q}(t, t')$. \textbf{Bottom panel}: The model (left) and unstructured (right) estimates of the multivariate functional random error covariance function $\textbf{S}(t, t')$.}
    \label{fig:covariance-reconstructions}
\end{figure}

Figure \ref{fig:covariance-reconstructions} displays filled-contour plots of the estimated multivariate covariance functions. The random-effects covariance function $\mathbf{Q}$ (top panel) is reconstructed by the model almost perfectly -- the model estimates (left panel) appear very similar to the unstructured estimates (right panel). The random-error covariance function $\mathbf{S}$ (bottom panel) is also well estimated, the model captures the general shape of the function. However, there are certain parts of $\mathbf{S}$ which the model cannot reconstruct. For example, in the knee-knee component of $\mathbf{S}$ in the region $t, t' \in [75, 100]$ there is a discrepancy between the model and unstructured estimates. We show in Appendix \ref{additional-random-effects} that this discrepancy is due to the diagonal assumption for $\mathbf{S}^*$ and can be resolved by allowing a small number of non-zero off-diagonal correlations in $\mathbf{S}^*$. Overall, however, we can conclude that the covariance functions are reconstructed well.

The functional ICC was estimated at $0.78$ (bootstrap $95\%$ CI = $[0.76, 0.81]$), indicating that approximately $78\%$ of the variability in the average hip and knee angle functions (after accounting for the fixed effects) can be explained by subject-level differences, and $22\%$ of variability is due to differences within subjects between sides. This high degree of ``clustering" highlights the presence of idiosyncratic running patterns that are consistent across both sides of the body.

\section{Discussion} \label{sec:bfmm-discussion}

This article has presented a multivariate functional mixed model for kinematic data from recreational runners collected during a treadmill run.
{Using an existing basis modelling approach \parencite{zhu_multivariate_2017}, we project the multivariate functional data onto a mv-FPCA basis to reduce it to a to a set of uncorrelated scores and fit a series of scalar linear mixed models to the scores.}
We have provided a frequentist implementation of the model which means it can be fitted using existing open-source software and adapted bootstrap and simulation techniques for inference on the fixed-effect function estimates.
{We introduced reconstructions and comparisons of the multivariate covariance structures to graphically assess the model, which demonstrated that the assumptions being made on these structures were appropriate for our data application.}
{We also extended the univariate functional ICC to summarise the degree of intra-subject correlation in our application, showing strong correlations in runner's bilateral kinematics (or equivalently, high inter-subject variability/ idiosyncrasy).}

{From a scientific perspective, we did not detect evidence of a statistically significant effect of retrospective injury status on the kinematic data after accounting for the other covariates.
It is difficult to directly compare this result with existing literature due to differences in study designs, and also because previous analyses have focused on discrete kinematic variables whereas we modelled kinematics over the course of the full running stride.
For example, the findings are consistent with the work of \textcite{messier_2-year_2018} who, in a large \emph{prospective} study of runner who were all asymptomatic at baseline, found that discrete measures of knee flexion were not significantly different between those who did and not become injured.
On the other hand, \textcite{bramah_is_2018} found a significant difference in knee flexion at the start of the stride between injured and uninjured runners in a retrospective study, were the injured runners \emph{were} symptomatic baseline.
Although our model did aim to comprehensively characterise effects of injury status on both hip and knee kinematics across the whole running stride, we cannot rule out the possibility that our injury groupings were too broad (both in terms of injury types/ location and the times since the runners were symptomatic), that our sample size was insufficient to detect a difference using these groupings, or that other joints or planes of motion are more important for RRIs.
}

{In contrast, we did find a strong, statistically significant effect of running speed on hip and knee kinematics -- participants who run faster tend to so by producing greater hip and knee flexion at various stages throughout the movement.
The functional effects that we have characterised using FDA are understood qualitatively in the biomechanical literature -- \textcite[p. 256]{grimshaw_bios_2007} explain that ``As speed increases, the flexion of hip and knee joints during the swing phase increases, this serves to reduce the moment of inertia of the limb, thus allowing for a faster swing. There may also be a slight increase in the degree of knee flexion at impact".
Additionally, almost identical qualitative effects of running speed were found by \textcite{orendurff_little_2018} in experiments where the individuals ran on a treadmill at multiple different speeds (i.e., speed was a within-subject variable).
However, their statistical modelling approach was limited to simply plotting the group average curves at the different speeds without any inference, and then reducing the curve data to discrete variables (e.g., peak knee flexion) and performing a repeated measures ANOVA on them, treating speed as a categorical covariate.
Although this might be sufficient in some applications, including speed as a continuous covariate in our multivariate functional mixed effects model allows us to obtain estimates that appropriately characterise the effect of the effect of speed across the whole running stride, obtain simultaneous inference on this effect that is valid across the whole running stride and across both the hip and knee joints, and to make predictions of full curves at different running speeds (Figure \ref{fig:speed-predictions}), while still accounting for repeated measures.
An added benefit of modelling the hip and knee jointly rather than fitting separate univariate models is that it leads to intuitive visualisations of combined effects on hip-knee kinematics using angle-angle diagrams (Figure \ref{fig:speed-predictions} (c)), which are an intuitive and practically useful tool for biomechanics researchers and practitioners conducting coordination research \parencite{lamb_assessing_2017}.
}

Some limitations and extensions of this work are as follows.
The kinematic data had already undergone filtering in the extraction step, as is typical for human movement data collected using motion capture systems, so further smoothing was not applied. 
However, in other scenarios where the data are less smooth, it may be desirable to regularise the estimated fixed effects functions.
This could be achieved by pre-smoothing the individual functional observations in the first-stage basis transformation or retaining fewer FPCs in the second stage.
However, in certain situations, heavily pre-smoothing individual observations may neglect uncertainty in their estimates in downstream analysis \parencite{bauer_introduction_2018}. 
The fixed effects estimates could also be post-smoothed by evaluating them on a grid and employing any scatterplot smoother \parencite{fan_two-step_2000, cui_fast_2022}. 
Finally, variable selection could be used in the scalar linear mixed effects models, which would lead to a sparse representation of the fixed-effect functions, i.e., each fixed-effect function would be represented by a small number of FPCs \parencite{morris_wavelet-based_2006, aston_linguistic_2010}.

In the second simulation scenario, and to a much lesser extent our scientific application, the random-effect and random-error covariance functions are reconstructed with error because the diagonal assumption for $\mathbf{Q}^*$ and $\mathbf{S}^*$ is too restrictive to fully capture the covariance structures. However, the approximation still works well to provide approximate fixed effects inference and summaries of the variance structure, i.e., the ICC. 
If better estimates of the covariance functions were required, a modification could be made to the current approach to allow a small number of off-diagonal elements in $\mathbf{Q}^*$ and $\mathbf{S}^*$ to be non-zero. We show in Appendix \ref{additional-random-effects} that unrestricted versions of $\mathbf{Q}^*$ and $\mathbf{S}^*$ can be estimated using the algorithm of \textcite{fieuws_pairwise_2006}, and variants of the graphical LASSO \parencite{friedman_sparse_2008} used to select which off-diagonal elements to retain. The final model could then be fitted with certain FPCs modelled in pairs or small groups, rather than completely independently.
{We worked with linear-time normalised and landmark-registered curves, but did not include the respective parameters of these transformations (i.e., curve lengths and landmark times) in subsequent analysis. It is likely that these parameters also depend on the covariates used in our model. A unified modelling of amplitude and phase \parencite[see, e.g.,][]{hadjipantelis_unifying_2015} could be achieved by modelling the phase variation parameters along with the mv-FPC scores, likely allowing for correlation among them. We leave further investigation of this approach to future work.}

Two main extensions of the model and application will be pursued. Our first goal is to extend the model to include all strides rather than an average for each side. As the strides admit a time ordering, \emph{longitudinal functional data analysis} methods will be required -- simply adding another level to the current model and ignoring the ordering of the strides may not be sufficient. There are a number of papers on univariate longitudinal functional data \parencite{greven_longitudinal_2010, park_longitudinal_2015, lee_bayesian_2019}, however we are developing bespoke methodology to handle the multivariate three-level case. The second extension is to include kinematic data from other joints, such as the ankle or pelvis, or from the other two planes of motion (i.e., frontal and transverse) in the model. This extension is more straightforward methodologically, but will be more computationally demanding and may provide interesting scientific results.

\section*{Acknowledgments}
This work was supported in part by Science Foundation Ireland (SFI) under grant numbers 18/CRT/6049 (EG), 19/FFP/7002 (SG, AJS and NB), and SFI/12/RC/2289\_P2 (RISC running dataset), and co-funded by the European Regional Development Fund.

\section*{Competing Interests}
The authors have no relevant financial or non-financial interests to disclose.

\section*{Implementation Details}
All analyses were performed in \proglang{R} version 4.1.2 \parencite{r_core_team_r_2022}. The \pkg{fda} package \parencite{ramsay_fda_2020} was used for the basis expansion and FPCA steps. The \pkg{lme4} package \parencite{bates_fitting_2015} was used to fit the univariate scalar mixed effects models. The \pkg{nlme} package \parencite{pinheiro_nlme_2022} was used to fit the bivariate scalar linear mixed models described in Appendix \ref{additional-random-effects}. The unstructured covariance functions were estimated using custom code that adapted the \pkg{denseFLMM} package \parencite{greven_denseflmm_2018}. The \pkg{mvtnorm} package \parencite{genz_mvtnorm_2021} was used to draw multivariate Gaussian samples for the simultaneous bands. The only computationally intensive part of the analysis was the bootstrap using 2500 bootstrap replicates, which took 15.3 minutes on 8 cores of a 2019 MacBook Pro with a 2.4 GHz Quad-Core Intel Core i5 processor and 8 GB of memory. 
We have prepared a \proglang{GitHub} repository containing custom functions to implement our methods and scripts to reproduce the results of the data analysis and simulations contained in the manuscript, which is available at \url{https://github.com/FAST-ULxNUIG/RISC1-fda-manuscript-01-code}.


\appendix
\renewcommand\thefigure{\thesection.\arabic{figure}}
\renewcommand\thefigure{\thesection.\arabic{table}}
\setcounter{figure}{0} 

\section{Unstructured Covariance Estimates} \label{unstructured-cov-est}

Unstructured estimates of $\mathbf{Q}$ and $\mathbf{S}$, the matrix-valued covariance functions associated with the bivariate functional random intercepts and the bivariate functional random error, respectively, can be obtained by extending the univariate multilevel FPCA method \parencite{di_multilevel_2009} to the multivariate case. To simplify this exposition, we slightly abuse notation and re-define $\mathbf{y}_{ij} (t)$ as a version of the bivariate functional response that has been centred around the fixed effects (i.e., $\boldbeta_0 (t) + \sum_{a=1}^A x_{ija}  \boldbeta_a (t)$ has been subtracted), so that
$$
\mathbf{y}_{ij} (t) =  \boldut + \boldepst,
$$
and consequently
$$\Expec [\mathbf{y}_{ij} (t) \  \mathbf{y}_{i'j'} (t') ^\top] = \delta_{ii'}\mathbf{Q} (t, t') + \delta_{ii'}\delta_{jj'}\mathbf{S} (t, t'),$$
where $\delta_{ii'} = 1$ if $i=i'$ and 0 otherwise. \textcite{di_multilevel_2009} proposed to use methods of moments (MoM) estimators for $\mathbf{Q}$ and $\mathbf{S}$.
\textcite{cederbaum_functional_2017} showed that these MoM estimators (and those for a more general class of FPC-based functional mixed models, such as the FPC-based functional random slope model of \textcite{greven_longitudinal_2010}) can be written as the solution to an ordinary least squares problem. \textcite{cederbaum_functional_2017} worked with univariate functional data observed on a regular grid; we adapt the methodology to multivariate functional data represented by a basis expansion.

We start with the basis representation for the centred data
$$
\mathbf{Y} (t) = \mathbf{Y}^* \boldsymbol{\Phi} (t),
$$
where $\mathbf{Y} (t) = (\mathbf{y}_{1 \text{,left}} (t) \ | \ \dots \ | \ \mathbf{y}_{N \text{,right}} (t))^\top$, $\mathbf{Y}^* = (\mathbf{y}_{1 \text{,left}}^* \ | \ \dots \ | \ \mathbf{y}_{N \text{,right}}^*)^\top$, $\mathbf{y}_{ij}^*$ is the vector of basis coefficients for the observation $\mathbf{y}_{ij}(t)$ and $\boldsymbol{\Phi}(t)$ is the multivariate basis matrix defined in Section \ref{sec:bfmm-first-stage-transform}. We can write the Kronecker product as
\begin{align}
    \mathbf{Y} (t) \otimes \mathbf{Y} (t') 
    &= (\mathbf{Y}^* \boldsymbol{\Phi} (t)) \otimes (\mathbf{Y}^* \boldsymbol{\Phi} (t')) \\
    &= (\mathbf{Y}^* \otimes \mathbf{Y}^*) (\boldsymbol{\Phi} (t) \otimes \boldsymbol{\Phi} (t')),
\end{align}
using the mixed-product rule $(\mathbf{A} \otimes \mathbf{B})(\mathbf{C} \otimes \mathbf{D}) = \mathbf{AC} \otimes \mathbf{BD}$ \parencite[Section 8.5.2, p. 135]{fieller_basics_2016}. Because $\mathbf{Y} (t) = \mathbf{ZU}(t) + \mathbf{E}(t)$, this is equivalent to
\begin{align}
     (\mathbf{ZU}(t) + \mathbf{E}(t)) \otimes (\mathbf{ZU}(t') + \mathbf{E}(t')) 
    &=   (\mathbf{ZU}(t) \otimes \mathbf{ZU}(t')) + \mathbf{E}(t) \otimes \mathbf{E}(t') \\
    & \ \ \ \ + \mathbf{ZU}(t) \otimes \mathbf{E}(t') + \mathbf{E}(t) \otimes \mathbf{ZU}(t') \\
    &=  (\mathbf{Z} \otimes \mathbf{Z})(\mathbf{U}(t) \otimes \mathbf{U}(t')) + \mathbf{E}(t) \otimes \mathbf{E}(t') \\
    & \ \ \ \ + \mathbf{ZU}(t) \otimes \mathbf{E}(t') + (\mathbb{I}_{2N} \otimes \mathbf{Z}) (\mathbf{E}(t) \otimes \mathbf{U}(t')),
\end{align}
using $(\mathbf{A}+\mathbf{B}) \otimes \mathbf{C} = \mathbf{A} \otimes \mathbf{C} + \mathbf{B} \otimes \mathbf{C}$ \parencite[Section 8.5.1, p. 135]{fieller_basics_2016} and then applying the mixed-product rule. 

We recognise that
$$
\mathbb{E}[\mathbf{U}(t) \otimes \mathbf{U}(t')] = \vect(\mathbb{I}_N) \vect(\mathbf{Q}(t, t')^\top)^\top,
$$
and
$$
\mathbb{E}[\mathbf{E}(t) \otimes \mathbf{E}(t')] = \vect(\mathbb{I}_{2N}) \vect(\mathbf{S}(t, t')^\top)^\top,
$$
because the rows of $\mathbf{U}(t)$ and $\mathbf{E}(t)$ each contain independent copies of the same process and
$$
\Expec[\mathbf{U}(t) \otimes \mathbf{E}(t')] = \Expec[\mathbf{E}(t) \otimes \mathbf{U}(t')] = \mathbf{0},
$$
because the entries of $\mathbf{U} (t)$ and $\mathbf{E}(t)$ are mutually uncorrelated. This allows us to write
\begin{align}
    \mathbb{E}[\mathbf{Y} (t) \otimes \mathbf{Y} (t')] &= (\mathbf{Z} \otimes \mathbf{Z}) \vect(\mathbb{I}_N) \vect(\mathbf{Q}(t, t')^\top)^\top + \vect(\mathbb{I}_{2N}) \vect(\mathbf{S}(t, t')^\top)^\top \\
    &= \vect(\mathbf{ZZ}^\top) \vect(\mathbf{Q}(t, t')^\top)^\top + \vect(\mathbb{I}_{2N}) \vect(\mathbf{S}(t, t')^\top)^\top.
\end{align}
with the second line using $\vect(\mathbf{AB}) = (\mathbf{B}^\top \otimes \mathbf{A}) \vect(\mathbf{I}_m)$ \parencite[Section 8.5.6, p. 138]{fieller_basics_2016}, if $\mathbf{A}$ is $m \times n$ and $\mathbf{B}$ is $n \times p$.

Finally, as the same basis is used to represent the matrix-valued covariance functions, we obtain
$$
\vect(\mathbf{Q}(t, t')^\top)^\top = \mathbf{q}^\top (\boldsymbol{\Phi} (t) \otimes \boldsymbol{\Phi} (t')),
$$
and 
$$
\vect(\mathbf{S}(t, t')^\top)^\top = \mathbf{s}^\top (\boldsymbol{\Phi} (t) \otimes \boldsymbol{\Phi} (t')),
$$
where $\mathbf{q}$ and $\mathbf{s}$ are $(K_{hip} + K_{knee})^2$-vectors of basis coefficients. This allows the relation to be written as
\begin{align}
    \mathbb{E}[(\mathbf{Y}^* \otimes \mathbf{Y}^*)] (\boldsymbol{\Phi} (t) \otimes \boldsymbol{\Phi} (t')) &= \vect(\mathbf{ZZ}^\top) \mathbf{q}^\top (\boldsymbol{\Phi} (t) \otimes \boldsymbol{\Phi} (t')) + \vect(\mathbb{I}_{2N}) \mathbf{s}^\top (\boldsymbol{\Phi} (t) \otimes \boldsymbol{\Phi} (t')) \\ \Longleftrightarrow \mathbb{E}[(\mathbf{Y}^* \otimes \mathbf{Y}^*)] &= \vect(\mathbf{ZZ}^\top)   \mathbf{q}^\top + \vect(\mathbb{I}_{2N})  \mathbf{s}^\top \\
    \Longleftrightarrow \mathbb{E}[(\mathbf{Y}^* \otimes \mathbf{Y}^*)] &= \underbrace{\Bigl[ \vect(\mathbf{ZZ}^\top) \Big| \vect(\mathbb{I}_{2N}) \Bigr]}_{\mathbf{X}}  \underbrace{\Bigl[ \mathbf{q} \Big|  \mathbf{s}\Bigr] ^\top}_{\boldsymbol{\beta}},
\end{align}
so that the basis coefficients can be obtained by solving the above least-squares problem. \textcite[Appendix B]{cederbaum_functional_2017} gives a computationally efficient representation of the least-squares solution based on rules for Kronecker products and $\vect$ operator notation; an associated implementation for univariate functional data observed on a grid is provided in the \pkg{denseFLMM} \proglang{R} package \parencite{greven_denseflmm_2018}. We adapt this software to calculate the unstructured covariance estimates for multivariate functional data represented by a basis expansion.

\renewcommand\thefigure{\thesection.\arabic{figure}}
\setcounter{figure}{0} 
\section{Additional Details on the ICC}\label{app:icc}
For univariate functional data (i.e., when $u_i(t)$ and $\varepsilon_{ij}(t)$ are scalars rather than vectors), \textcite{di_multilevel_2009} introduced the functional ICC
$$
\text{ICC} = \frac{\int_{\mathcal{I}} \Var\{u_i (t)\} \mathrm{d}t}{\int_{\mathcal{I}} \Var\{u_i (t)\} \mathrm{d}t + \int_{\mathcal{I}} \Var\{\varepsilon_{ij} (t)\} \mathrm{d}t},
$$
where $\mathcal{I} = [0, T]$. The natural extension of this quantity for multivariate functional data is obtained by summing over the dimensions of the multivariate function. Letting $\mathcal{P} = \{\text{hip}, \text{knee}\}$, we define the ICC for our model as
\begin{align*}
    \text{ICC} &= \frac{\sum_{p \in \mathcal{P}} \int_{\mathcal{I}} \Var\{u_i^{(p)} (t)\} \mathrm{d}t}{\sum_{p \in \mathcal{P}} \int_{\mathcal{I}} \Var\{u_i^{(p)} (t)\} \mathrm{d}t + \sum_{p \in \mathcal{P}} \int_{\mathcal{I}} \Var\{\varepsilon_{ij}^{(p)} (t)\} \mathrm{d}t}.
\end{align*}
Now, we can rewrite
\begin{align*}
    \sum_{p \in \mathcal{P}} \int_{\mathcal{I}} \Var\{u_i^{(p)} (t)\} \mathrm{d}t
&=
\sum_{p \in \mathcal{P}} \int_{\mathcal{I}} \Var\{\sum_{k=1}^{\widetilde{K}} u_{ik}^* \psi^{(p)}_k(t)\} \mathrm{d}t \\
&= 
\sum_{k=1}^{\widetilde{K}} \sum_{k'=1}^{\widetilde{K}} \underbrace{\Cov(u_{ik}^*, u_{ik'}^*)}_{= q_{k}^{} \delta_{kk'}}
\underbrace{
\sum_{p \in \mathcal{P}} \int_{\mathcal{I}} \{\psi^{(p)}_k(t) \psi^{(p)}_{k'}(t)\} \mathrm{d}t}_{= \delta_{kk'}} =
\sum_{k=1}^{\widetilde{K}} q_k,
\end{align*}
where $\delta_{kk'} = 1$ if $k=k'$ and 0 otherwise. Similarly, we have $\sum_{p \in \mathcal{P}} \int_{\mathcal{I}} \Var\{\varepsilon_{ij}^{(p)} (t)\} \mathrm{d}t = \sum_{k=1}^{\widetilde{K}} s_k$, which means that the ICC can be written as
$$
\ICC = \frac{\sum_{k=1}^{\widetilde{K}} q_k}{
\sum_{k=1}^{\widetilde{K}} q_k + \sum_{k=1}^{\widetilde{K}} s_k}.
$$

\clearpage
\setcounter{figure}{0} 
\section{Simulation Study} \label{sec:bfmm-simulation}

We perform a short simulation study to demonstrate the properties of our method in two scenarios. The first scenario assumes that the fixed effects, random effects and random error terms are generated by the same bivariate FPC basis, which means our proposed model is correctly specified. In the second scenario, a different bivariate basis is used for the fixed effects, random effects and random error terms, in which case our proposed model is an approximation.

To liken the simulation setting to the real data application, we use empirical parameter estimates to generate the data. We generate data from $N = 280$ subjects with two bivariate functional observations each to mirror our study design, i.e., hip and knee angles measured on both sides of the body for each subject. For each subject, we generate two covariates -- a continuous self-selected running speed variable and a binary sex variable; both covariates were included in our final model for the real data analysis. Observations are simulated according to the model
$$
\mathbf{y}_{ij} (t) = \boldsymbol{\beta}_0 (t) + \boldbeta_1 (t) \times\text{sex}_i + \boldbeta_2 (t) \times \text{speed}_i + \boldut + \boldepst,
$$
for $i = 1,\dots, 280, \ j = 1, 2, \ t \in [0, 100]$ and $\text{sex}_i \in \{0, 1\}$. This is a version of model \eqref{FMM} with $A = 2$, we have simplified the model to have only two fixed effect parameters to keep the simulation and presentation of results concise.

\begin{figure}[!htbp]
    \centering
    \includegraphics[width = 1\textwidth,page=2]{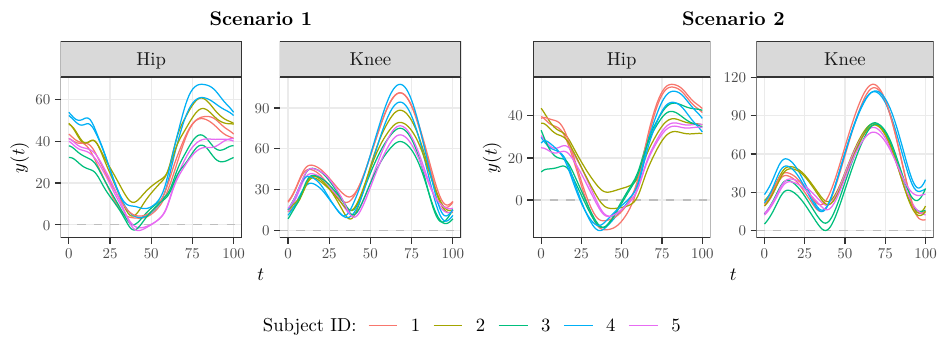}
    \caption{10 simulated bivariate functional observations from the first and second simulation scenarios.}
    \label{fig:simulated-data}
\end{figure}

In both scenarios, we use empirical estimates for the fixed effects functions $\boldbeta_0 (t)$, $\boldbeta_1 (t)$ and $\boldbeta_2 (t)$, and use 13 bivariate basis functions to generate the model terms because 13 bivariate FPCs explained $99\%$ of the variability in the real data application. For the first scenario, we use the same basis of 13 empirical bivariate FPCs to generate $\boldut$ and $\boldepst$. In the second scenario, we use 13 bivariate Fourier basis functions for $\boldut$ and 13 bivariate polynomial basis functions for $\boldepst$ \parencite{happ-kurz_object-oriented_2020}. In both scenarios, we draw the basis coefficients of the functional random intercepts and functional random errors from a multivariate Gaussian distribution with a diagonal covariance matrix, and use empirical estimates for the coefficient variances (i.e., $\widehat{q}_k$ and $\widehat{s}_k$, $k=1, \dots 13$ from our fitted model). Figure \ref{fig:simulated-data} shows 10 simulated bivariate functional observations from both data-generating scenarios.

We evaluate each fixed effect estimate in terms of integrated squared error (ISE), defined as 
$$
\text{ISE}(\widehat{\boldsymbol{\beta}}_a) = \int_0^{100} \left\{\widehat{\beta}^{(hip)}_a (t) - \beta^{(hip)}_a (t)\right\}^2 \mathrm{d}t + 
\int_0^{100} \left\{\widehat{\beta}^{(knee)}_a (t) - \beta^{(knee)}_a (t)\right\}^2 \mathrm{d}t,
$$
for $a = 0, 1, 2$. We consider the coverage probabilities of the pointwise confidence intervals and simultaneous confidence bands. The matrix-valued covariances of the bivariate functional random intercepts and bivariate functional random errors are also assessed in terms of ISE
$$
\text{ISE}(\widehat{\mathbf{Q}}) = \int_0^{100}  \int_0^{100}  \sum_{p \in \{\text{hip, knee}\}} \sum_{p' \in \{\text{hip, knee}\}} \left\{\widehat{Q}^{(pp')}(t, t') - Q^{(pp')}(t, t')\right\}^2 \mathrm{d} t \mathrm{d}t',
$$
with the $\text{ISE}$ for $\widehat{\mathbf{S}}$ defined similarly. The ICC and the number of bivariate FPCs retained are also examined. For both scenarios, we generate 500 simulated datasets. The mv-FPCA truncation is fixed at $99.99\%$, $1000$ bootstrap replications are used, \num{10000} multivariate Gaussian samples are drawn for the confidence bands and a nominal level of $95\%$ is used for the confidence intervals.

\begin{figure}[!htbp]
    \centering
    \includegraphics[width = 1\textwidth,page=3]{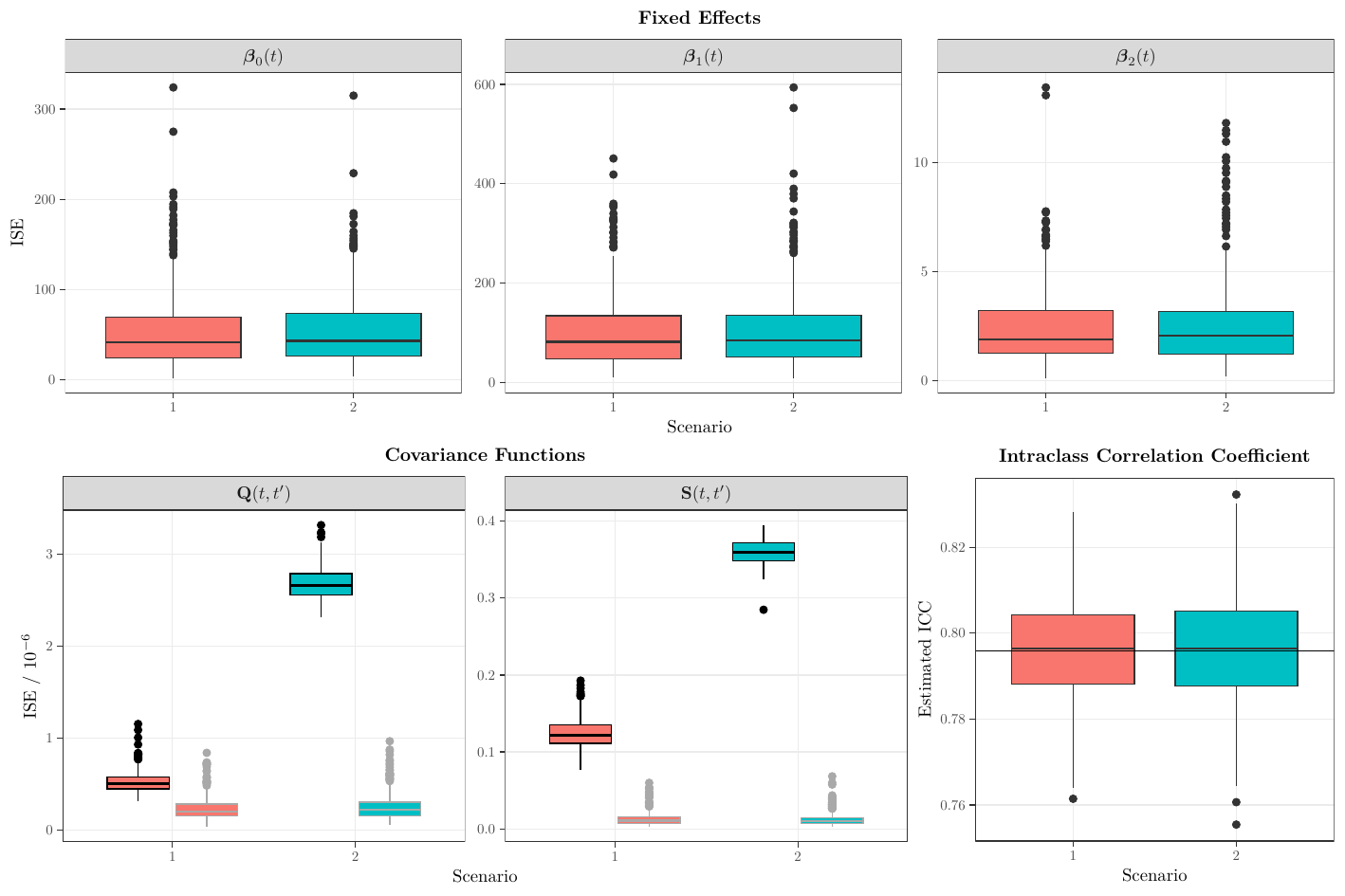}
    \caption{Estimation results from the simulation study under the two data-generating scenarios (1 = red, 2 = turquoise). \textbf{Top panel}: The ISE values for the fixed effects functions $\boldsymbol{\beta}_1 (t)$ (left panel) and $\boldsymbol{\beta}_2 (t)$ (right panel). \textbf{Middle panel}: The ISE values for the random-effects covariance function $\mathbf{Q}(t, t')$ (left panel) and the random-error covariance function $\mathbf{S}(t, t')$ (right panel). The boxplots with black trimming show the model estimates and are presented next to unstructured estimates shown by the boxes with grey trimming. \textbf{Bottom panel}: The ICC estimates with the true ICC value overlaid as a solid horizontal black line.}
    \label{fig:simulation-results}
\end{figure}

Figure \ref{fig:simulation-results} shows the estimation results. There is little difference in fixed effects estimation between the two scenarios (top panel). The ISE values for $\boldsymbol{\beta}_1 (t)$ are larger than those for $\boldsymbol{\beta}_2 (t)$, simply because of a smaller signal-to-noise ratio for this effect. For the random-effect and random-error covariances (middle panel), estimation is substantially better under Scenario 1 than under Scenario 2. The model estimates (boxes with a black outline) are much closer to the unstructured estimates, (boxes with a grey outline) under Scenario 1. The ICC is estimated similarly and centred on the true value in both scenarios (bottom panel). To summarise, the fixed effects functions and ICC are estimated comparably well in both scenarios. However, when the model is incorrectly specified in Scenario 2 (i.e., different bases are used for the fixed effect, random-intercept and random-error functions), estimation quality of the covariance structures is reduced significantly. As shown empirically in our data application in Section \ref{sec:bfmm-results}, this is because of the restriction of the diagonal assumption for $\mathbf{Q}^*$ and $\mathbf{S}^*$ -- correlation (in the random-intercept and random-error terms) between FPCs is required to fully reconstruct the covariance functions.

Table \ref{tab:coverage-probs} contains the coverage probability estimates from the simulation along with Monte Carlo standard errors, which quantify simulation uncertainty due to the finite number of simulation repetitions \parencite{morris_using_2019}. Pointwise and simultaneous coverage probabilities are close to nominal and performance of the bootstrap and Wald methods is comparable in both scenarios. This indicates that although the covariance structures are reconstructed with error in Scenario 2, they still allow for fast approximate uncertainty estimates that work reasonably well to be constructed. It should be noted that the reported pointwise coverage is an ``across the function" average -- it is averaged across the functional domain and across the two dimensions of the bivariate function. Next, we demonstrate that the Wald intervals tend to over-cover and under-cover at different points along the function in Scenario 2 and that, in comparison, the pointwise coverage of the bootstrap intervals is stable and may be preferred.


{
\begin{table}[]
\centering
\begin{tabular}{cccccc}
\hline
\multirow{2}{*}{\textbf{Method}} &
  \multirow{2}{*}{\textbf{\begin{tabular}[c]{@{}c@{}}Coverage\\ Type\end{tabular}}} &
  \multirow{2}{*}{\textbf{Scenario}} &
  \multicolumn{3}{c}{\textbf{Parameter}} \\ \cline{4-6} 
                           &                               &   & $\boldsymbol{\beta}_0(t)$ & $\boldsymbol{\beta}_1(t)$ &
                           $\boldsymbol{\beta}_2(t)$ \\ \hline
\multirow{4}{*}{Wald}      & \multirow{2}{*}{Pointwise}    & 1 & 0.95 (0.01)  & 0.96 (0.01)  & 0.96 (0.01) \\
                           &                               & 2 & 0.94 (0.01)  & 0.95 (0.01)  & 0.95 (0.01)  \\ \cline{2-3}
                           & \multirow{2}{*}{Simultaneous} & 1 & 0.94 (0.01)  & 0.96 (0.01)  & 0.96 (0.01)  \\
                           &                               & 2 & 0.92 (0.01)  & 0.95 (0.01)  & 0.94 (0.01)  \\ \cline{1-3}
\multirow{4}{*}{Bootstrap} & \multirow{2}{*}{Pointwise}    & 1 & 0.95 (0.01)  & 0.95 (0.01)  & 0.95 (0.01)  \\
                           &                               & 2 & 0.95 (0.01)  & 0.95 (0.01)  & 0.95 (0.01)  \\ \cline{2-3}
                           & \multirow{2}{*}{Simultaneous} & 1 & 0.92 (0.01)  & 0.95 (0.01)  & 0.94 (0.01)  \\
                           &                               & 2 & 0.93 (0.01)  & 0.96 (0.01)  & 0.94 (0.01)  \\ \hline
\end{tabular}
\caption{Coverage probability estimates of the $95\%$ pointwise confidence intervals and simultaneous confidence bands obtained from the simulation. Estimated Monte Carlo standard errors are shown in brackets to quantify simulation uncertainty. The pointwise coverage probabilities have been averaged across the functional domain and across dimensions of the bivariate function.}
\label{tab:coverage-probs}
\end{table}
}

\begin{figure}[h]
    \centering
    \includegraphics[width = 1\textwidth,page=4]{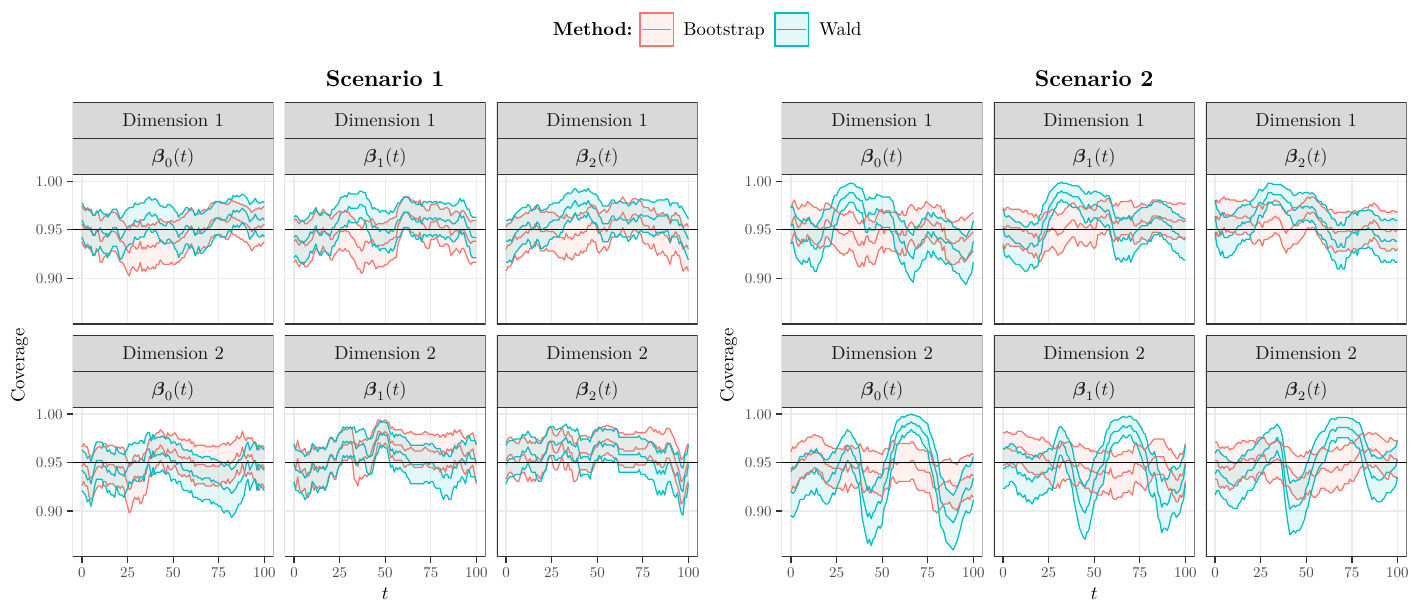}
    \caption{Coverage probability estimates of the $95\%$ pointwise confidence intervals from the simulation in Section \ref{sec:bfmm-simulation}. The central line represents the coverage probability estimate and the ribbon represents a $95\%$ pointwise confidence interval which accounts for uncertainty in the simulation. The nominal coverage level of $95\%$ is indicated by the solid black line.}
    \label{fig:simulation-coverage}
\end{figure}

Figure \ref{fig:simulation-coverage} shows the pointwise coverage probability estimates of the $95\%$ pointwise confidence intervals from the simulation in Section \ref{sec:bfmm-simulation} for both the bootstrap (red) and Wald (turquoise) approaches. The coverage probability estimates are accompanied by ribbons representing $95\%$ pointwise confidence intervals, accounting for uncertainty in the simulation. For Scenario 1, the pointwise coverage probability of both approaches appears relatively stable across the functional domain and across the dimensions of the bivariate function. For Scenario 2, the bootstrap coverage probabilities appear stable, however the Wald intervals appear to over-cover and under-cover at different points along the function. For this scenario, coverage of the Wald intervals ranges between $0.89$ and $0.99$, whereas coverage of the bootstrap intervals ranges between $0.92$ and $0.97$. The varying coverage of the Wald intervals is likely caused by the worse estimation of the random effect and random error covariance functions in this scenario.

\clearpage
\setcounter{figure}{0} 
\section{Additional Results} \label{additional-results}
\subsection{Basis Transformation}  \label{additional-bfmm-basis-transformation}
\begin{figure}[h]
    \centering
    \includegraphics[width = 0.99\textwidth,page=5]{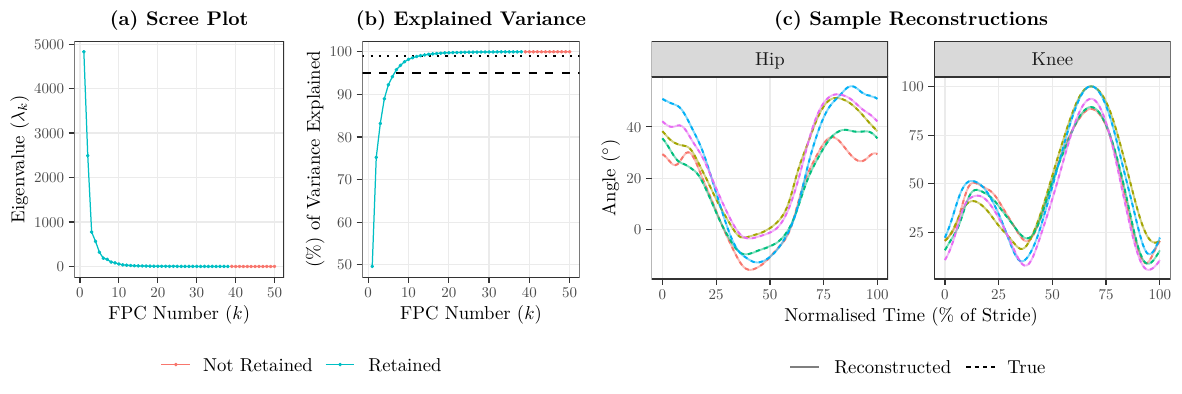}
    \caption{Results for the second-stage basis transformation using bivariate FPCA. \textbf{(a)} Plot of the first 50 eigenvalues of the bivariate FPCA. The FPCs retained for the analysis are shown in turquoise and those not retained are shown in red. \textbf{(b)} The cumulative percentage of variance explained by the bivariate FPCs. The dashed and dotted black horizontal lines represent the $95\%$ and $99\%$ variance-explained thresholds, respectively. Again, the FPCs retained for the analysis are shown in turquoise and those not retained are shown in red. \textbf{(c)} The reconstruction given by the bivariate FPCA for a random sample of $5$ functional observations in the data. The ``true" functions, taken to be the data before the bivariate FPCA, are shown by a light solid line. The bivariate FPCA reconstructions are overlaid in a darker dashed line.}
    \label{fig:mfmm-basis-transformation}
\end{figure}

Figure \ref{fig:mfmm-basis-transformation} shows results of the second-stage basis transformation, i.e., the bivariate FPCA. The scree plot in panel (a) shows that the eigenvalues decrease rapidly. This is reflected in the plot in panel (b), which shows the cumulative percentage of variance explained by each successive FPC. It is clear that the first few FPCs explain a large amount of the variance in the data. The dashed and dotted horizontal lines represent cumulative variance explained thresholds of $95\%$ and $99\%$ respectively. The plot indicates that although $\widetilde{K} = 38$ FPCs were retained, a similar reconstruction may have been achieved with $\widetilde{K} \approx 25$. Panel (c) shows the reconstructions of a random sample of five functional observations. Here, the ``truth" is the functional observation before the bivariate FPCA. As expected, the functions are reconstructed almost perfectly.

\subsection{Random Effects} \label{additional-random-effects}

We restrict our attention to the reconstruction of $\mathbf{S}$ because it appears less well-estimated than $\mathbf{Q}$.
Because the FPCA step is near-lossless, it must be the diagonal restriction for $\mathbf{S}^*$ rather than a truncation effect that is limiting the reconstruction of $\mathbf{S}$. Estimating a multivariate linear mixed model with unstructured $\mathbf{S}^*$ would involve jointly estimating $\widetilde{K} (\widetilde{K} + 1) / 2~=~741$ parameters, which is not computationally feasible.
We therefore use the pairwise modelling approach of \textcite{fieuws_pairwise_2006} to estimate the full multivariate model by modelling each pair of outcomes (in our case, FPC scores) separately and combining the estimates. 

\begin{figure}[h]
    \centering
    \includegraphics[width = 0.8\textwidth,page=7]{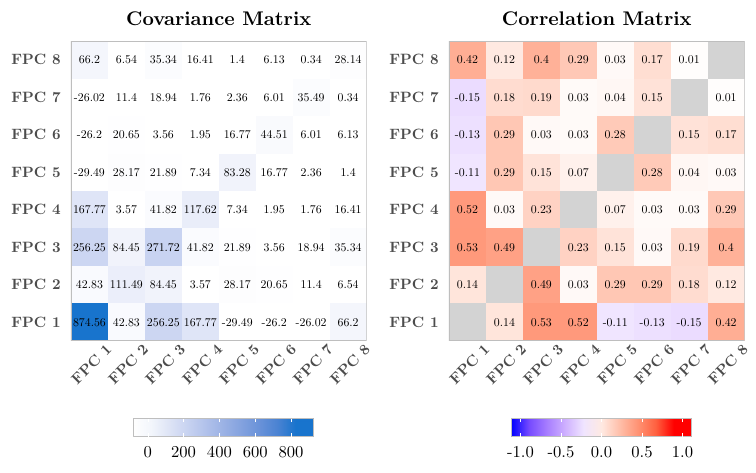}
    \caption{\textbf{Left panel}: The unstructured estimate of $\textbf{S}^*$ for the first 8 FPCs. \textbf{Right panel}: The corresponding correlation matrix.}
    \label{fig:cov-and-cor-mat}
\end{figure}

Figure \ref{fig:cov-and-cor-mat} displays the unstructured estimate of $\textbf{S}^*$ (left panel) and its corresponding correlation matrix (right panel) for for the first 8 FPCs. Some off-diagonal elements of $\textbf{S}^*$ are estimated at reasonably large values and correspond to moderate correlations, e.g., between FPC1 and FPC4, and between FPC1 and FPC3. This can be understood as a Simpson's Paradox type phenomenon -- the FPC scores are marginally uncorrelated but become correlated when centred around the subject's average. Next, we show that allowing for a small number of these correlations improves the reconstruction of $\textbf{S}$.

\begin{figure}[h]
    \centering
    \includegraphics[width = 0.75\textwidth,page=8]{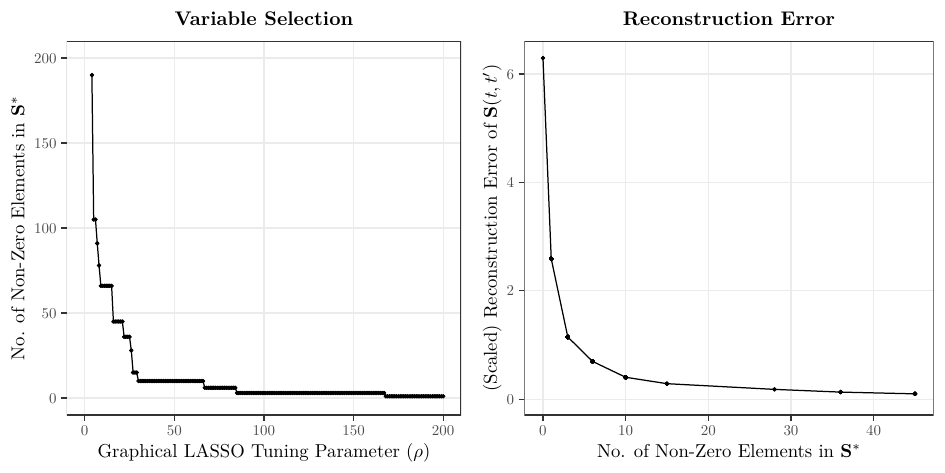}
    \caption{\textbf{Left panel}: The number of off-diagonal elements of $\mathbf{S}^*$ estimated as non-zero by the graphical LASSO for a range of values of the tuning parameter. The $y$-axis range is limited to $[0,200]$ to aid interpretation. \textbf{Right panel}: The reconstruction error of $\mathbf{S}$ using a solution of $\mathbf{S}^*$ with different numbers of non-zero off-diagonal elements as identified by the graphical LASSO. The $x$-axis range is limited to $[0,45]$ to aid interpretation.}
    \label{fig:glasso-results}
\end{figure}

As an exploratory exercise, we apply the graphical LASSO \parencite{friedman_sparse_2008} to the unrestricted $\mathbf{S}^*$ to select off-diagonal elements to retain. The left panel of Figure \ref{fig:glasso-results} shows the number of non-zero off-diagonal elements in the solution of $\mathbf{S}^*$ for a range of values for the tuning parameter, which controls the weighting of the $\ell_1$ penalty on the off-diagonal elements of the inverse $\textbf{S}^{*-1}$. For very large values of the tuning parameter, sparse solutions of $\mathbf{S}^*$ with three and one off-diagonal elements are returned. The solution with three off-diagonal elements allows all pairwise correlations between FPC1, FPC3 and FPC4 to be non-zero and appears to be reasonably consistent with the structure in Figure \ref{fig:cov-and-cor-mat}.

\begin{figure}[h]
    \centering
    \includegraphics[width = 1\textwidth,page=9]{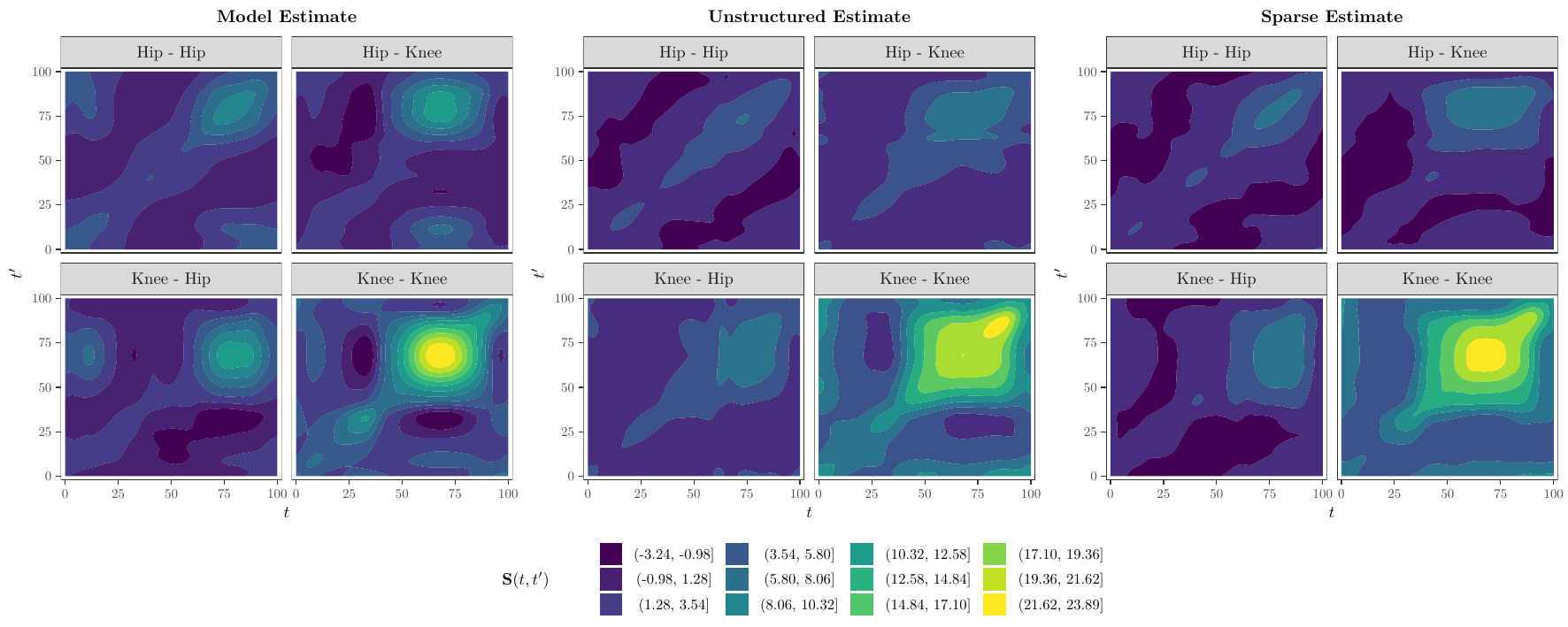}
    \caption{The left and middle panels show the model and unstructured estimates of the multivariate functional random error covariance function $\textbf{S}$, respectively, the same as in Figure \ref{fig:covariance-reconstructions}. The right panel shows a reconstruction of $\textbf{S}$ using a sparse version of $\mathbf{S}^*$, in which three off-diagonal elements are allowed to be non-zero. The three off-diagonal elements were selected by the graphical LASSO and correspond to the covariance terms between FPC1, FPC3 and FPC4.}
    \label{fig:sparse-reconstruction}
\end{figure}

For each solution, we then reconstructed $\mathbf{S}$ and assessed the integrated squared reconstruction error. The graphical LASSO was only used to select \emph{which} elements to set to zero -- the estimates of the non-zero coefficients from the original fit, rather than the penalised estimates from the graphical LASSO, were used in the reconstruction. The right panel of Figure \ref{fig:glasso-results} shows the reconstruction error for solutions with differing numbers of off-diagonal elements. It can be seen that quite substantial reductions in the reconstruction error are achieved by allowing one and three off-diagonal elements in $\mathbf{S}^*$ to be non-zero. Figure \ref{fig:sparse-reconstruction} contains the same reconstructions of $\mathbf{S}$ as in Figure \ref{fig:covariance-reconstructions}, except that they are also accompanied by a ``sparse" reconstruction (right panel), which uses a version of $\textbf{S}^*$ with three off-diagonal elements identified by the graphical LASSO allowed to be non-zero. Allowing the off-diagonal elements to be non-zero helps to improve the reconstruction, particularly in the knee-knee component in the region $t_1, t_2 \in [75, 100]$. This is, however, an exploratory investigation which identifies limitations of the model fitted to empirical data, and would need to be verified in more general scenarios before being used more broadly as a method.

\clearpage

\clearpage
\setcounter{figure}{0} 
\setcounter{table}{0}
\section{Comparison with Existing Methods}\label{app:comparisons}
In this section, we present the results of applying other approaches to our dataset. It is important to emphasise that we are not trying to prove superiority of any one approach. Instead, we are testing if and how state-of-the-art methods perform in realistic scenarios, as advocated by \textcite{sergazinov_case_2023}. Understanding the computational effort involved with different methods is valuable when considering scaling analysis up to larger and more complex datasets. In addition, the comparison enables us to understand the influence of different modelling strategies and assumptions on our real application.
The two alternative approaches that we apply are the multiFAMM, which is currently (to the best of our knowledge) the only publicly-available \proglang{R} package for fitting multivariate functional mixed effects models, and the fast univariate inference (FUI) method, which is a current state-of-the-art approach for fitting univariate functional mixed effects models to large datasets.

\subsection{multiFAMM}
The multiFAMM is implemented in the \proglang{R} package \pkg{multifamm} \parencite{volkmann_multifamm_2021}. We re-iterate that this is not a direct comparison with our proposed model, as the multiFAMM method is designed for very general settings, whereas our approach makes assumptions that are tailored to the smooth kinematic data in our application. For example, the covariance smoothing in multiFAMM for sparse and irregularly observed functions could be replaced by the approaches described in Appendix \ref{unstructured-cov-est} for covariance estimation or by modern developments for fast ml-FPCA \parencite{cui_fast_2023}. However, we proceed with the default implementation now.

Choices are required regarding the parameters used to fit the multiFAMM model. The fixed effects are modelled using P-splines, so the number of B-spline basis functions needs to be chosen. As we evaluate the functional data on $101$ equidistant points for analysis, we choose $K=26$ B-spline basis functions for each of the regression coefficient functions, based on the recommendation of $\min$\{no. of sampling points $\times \allowbreak \frac{1}{4} \allowbreak, 35$\} given by \textcite[p. 126]{ruppert_semiparametric_2003}. For the number of marginal basis functions used to smooth the univariate covariance surfaces within each dimension, using $K=26$ is not computationally feasible on a laptop with 8GB of RAM. However, the default value of $K=5$ in the \texttt{multiFAMM()} function is likely too small for the application at hand. Therefore, we trial the values $5, 8, 10$ and $15$. We also trial values of $0.9$ and $0.95$ for the proportion of variance explained (PVE) cutoff which is used to choose the number of mv-FPCs retained at each level. 

\begin{table}[ht]
\centering
\begin{tabular}{rrr}
  \toprule
{\bfseries $K$ Marginal} & {\bfseries PVE} & {\bfseries Time (mins)} \\ 
  \midrule
5 & 0.90 & 10.86 \\ 
  8 & 0.90 & 23.49 \\ 
  10 & 0.90 & 48.77 \\ 
  15 & 0.90 & 99.23 \\ 
  5 & 0.95 & 19.06 \\ 
  8 & 0.95 & 33.82 \\ 
  10 & 0.95 & 53.78 \\ 
  15 & 0.95 & 128.61 \\ 
   \bottomrule
\end{tabular}
\caption{Computation time for the multiFAMM model with different settings.} 
\label{tab:mfamm-comp-time}
\end{table}

Table \ref{tab:mfamm-comp-time} displays the computation times for the multiFAMM in the different settings. The default setting in the software ($K=5$ and $\text{PVE}=0.95$) took $19.06$ minutes. As the number of marginal basis functions is increased, the computation time increases greatly. This is in a large part due to the increased overhead involved in the two-dimensional covariance smoothing. However, we also noticed that the final scalar additive mixed model fit took longer as $K$ was increased. We hypothesise that this is because more complex covariance structures were estimated using a more flexible basis, requiring more mv-FPCs to explain and hence more parameters to estimate in the final model. 
This can also be inferred by comparing the computation times between the $\text{PVE}=0.9$ and $\text{PVE}=0.95$ settings for a given $K$. Bearing in mind that the reduced dataset is $<100$ times the size of the full dataset, some of these computation times are substantial and potentially prohibitive. However, they do appear to be sensitive to the settings used, with times ranging between just under $11$ minutes and just over $2$ hours and $8$ minutes.

\begin{figure}
    \centering
    \includegraphics[width = 0.75\textwidth]{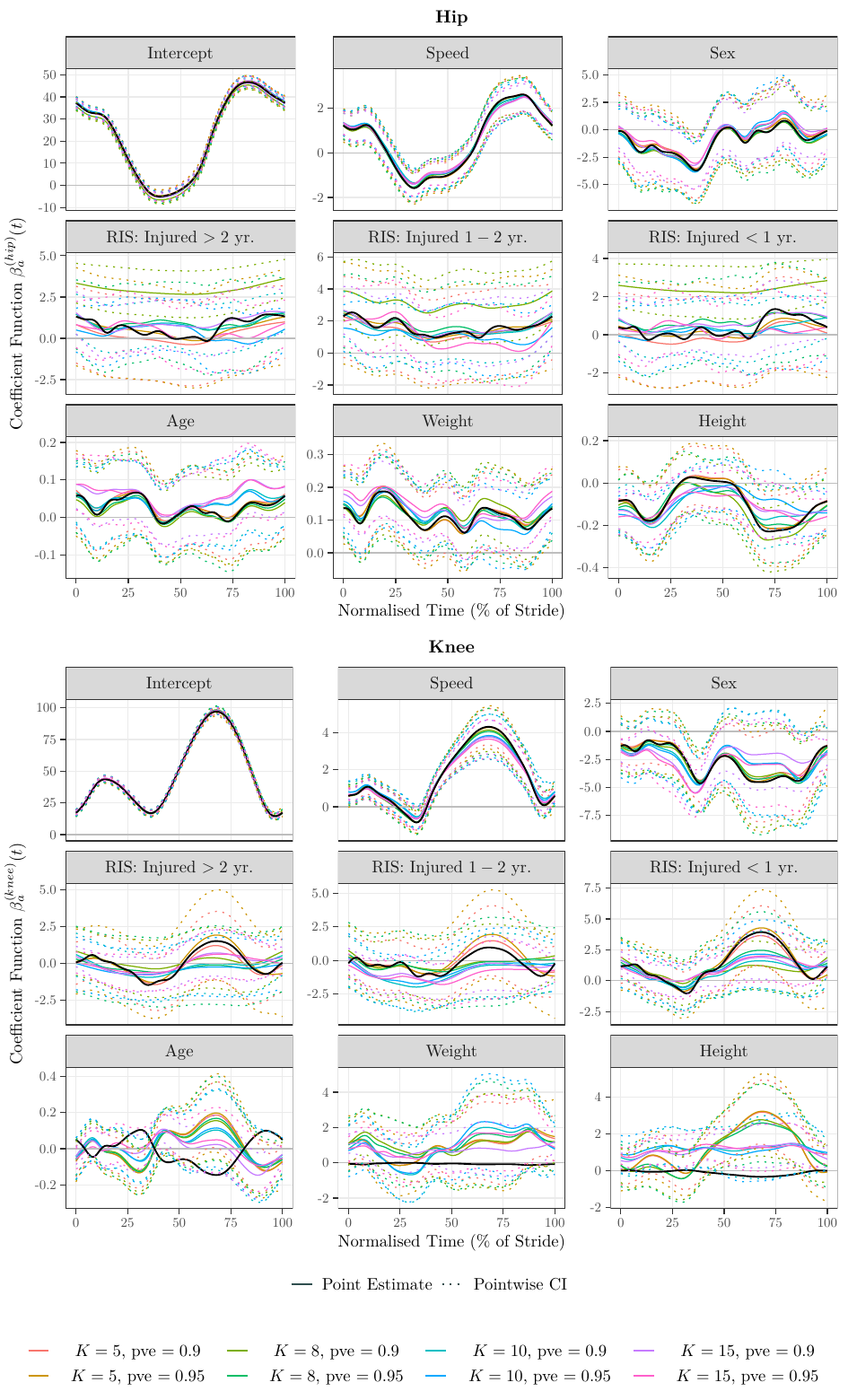}
    \caption{Comparison between the results of our proposed model and the results of applying the multiFAMM model with different settings. The solid coloured lines indicate the point estimates from the multiFAMM model with different settings. The black solid lines indicate the point estimates from our proposed model. The dotted coloured lines represent pointwise $95\%$ confidence intervals from the multiFAMM model.}
    \label{fig:multifamm-comparison}
\end{figure}

Figure \ref{fig:multifamm-comparison} displays the results of the comparison with the multiFAMM. The multiFAMM point estimates are displayed as coloured lines and the estimates from our proposed model are overlaid as solid black lines. The multiFAMM does not readily produce simultaneous confidence bands and it is computationally infeasible to obtain them via a bootstrap. Therefore, we present pointwise confidence intervals, which are indicated as dotted coloured lines. Some of the multiFAMM estimates also appear to be sensitive to the settings, as evidenced by the differences between the different coloured lines. 
The most obvious discrepancy is for the RIS coefficients (middle panel) for the hip angle at $K=8$ and $\text{PVE}=0.9$, which appear to be very different to the estimates from the other settings and lie outside some of their confidence intervals. 
Apart from this and despite discrepancies between the settings, the point estimates of the hip angle coefficients appear to be roughly centred on and in agreement with the estimates from our proposed model.
For the knee angle, the intercept, speed and sex coefficient estimates from the multiFAMM are reasonably stable across the settings and in agreement with the estimates from our proposed model.
For the RIS coefficients, flatter estimates are obtained when $K$ is increased, which likely reflects increased smoothing of these effects as the variance structure in the scalar additive mixed model changes.

\begin{figure}
    \centering
    \includegraphics[width = 0.75\textwidth]{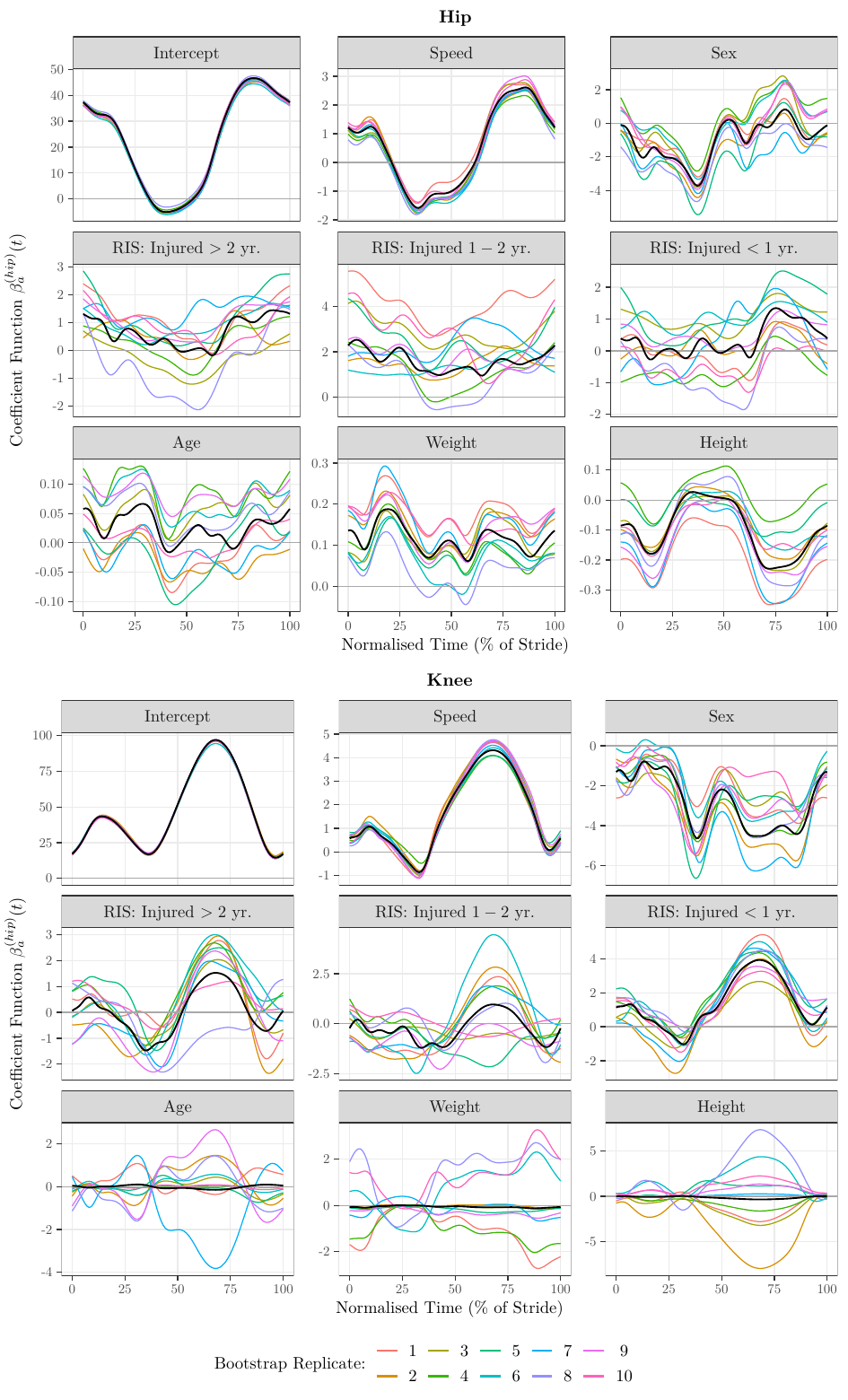}
    \caption{Results of a bootstrap of subjects for the multiFAMM. $B=10$ bootstrap replicate datasets were produced and each time a multiFAMM was fitted using $K=5$ marginal basis functions to smooth the univariate covariance functions and $\text{PVE}=0.95$. The coloured lines indicate the multiFAMM point estimates and the black lines indicate the point estimates from our proposed model.}
    \label{fig:multifamm-comparison-bootstrap}
\end{figure}

Only the multiFAMM estimates of age, weight and sex for the knee angle disagree with the estimates from our proposed model. Our original analysis found these effects to be small and not statistically significantly different from $0$.
The pointwise confidence intervals for the multiFAMM estimates of these effects also mostly contain $0$, but the shape and magnitude of the regression coefficient functions are different.
To qualitatively examine the stability of these coefficient estimates, we conducted a bootstrap of subjects using the default settings ($K=5$ and $\text{PVE}=0.95$) and $B=10$ bootstrap replicates. In total, the $10$ bootstrap model fits took $3.03$ hours to complete. Figure \ref{fig:multifamm-comparison-bootstrap} shows the results of the bootstrap. The point estimates from each bootstrap replicate are shown as coloured lines and the point estimates from our proposed model are overlaid as black solid lines. The direction and magnitude of the aforementioned effects of age, weight and height for the knee vary substantially across bootstrap replicates, highlighting that these estimates are unstable. This may be the result of the large number of parameters being estimated jointly in the scalar additive mixed model or variability in the estimated mv-FPCs being used as basis functions. Further investigations could use simulated data based on this setting. In general, however, scientific conclusions from the main text remain the same.

\subsection{Fast Univariate Inference (FUI)}
We compare point estimates and pointwise confidence intervals for the fixed effects with the FUI method \parencite{cui_fast_2022}. 
FUI is designed for univariate functional data, so it does not estimate the multivariate covariance structure or provide joint confidence bands for multivariate regression coefficient functions. Therefore, we fit a separate FUI model to the knee and the hip dimensions and compare fixed-effect point estimates and pointwise confidence intervals. FUI admits two ways to construct pointwise confidence intervals for Gaussian functional mixed models -- bootstrap and analytic approaches \parencite{cui_fast_2022}. Naturally, we compare the FUI bootstrap intervals to our bootstrap intervals and the analytic intervals to our Wald intervals. FUI point estimates are obtained by fitting univariate scalar linear mixed effects models at each sampling point and are hence the same regardless of what approach is used for inference. 
To fit the FUI, we discretise the functional observations on a grid of $101$ equidistant points $0,1, \dots, 100$. For the bootstrap approach, we use $B=100$ bootstrap replicates (this is fixed as the default in the software implementation \parencite[Supplementary Material]{cui_fast_2022}) and parallelise model fits across $8$ cores.

\begin{table}[ht]
\centering
\begin{tabular}{llr}
  \toprule
{\bfseries Approach} & {\bfseries Dimension} & {\bfseries Time (secs)} \\ 
  \midrule
Analytic & Hip & 72.33 \\ 
  Analytic & Knee & 72.00 \\ 
  Bootstrap & Hip & 167.73 \\ 
  Bootstrap & Knee & 153.05 \\ 
   \bottomrule
\end{tabular}
\caption{Computation time for fitting separate FUI models to the hip and knee data using bootstrap and analytic approaches.} 
\label{tab:fui-comp-time}
\end{table}

\begin{figure}
    \centering
    \includegraphics[width = 0.8\textwidth]{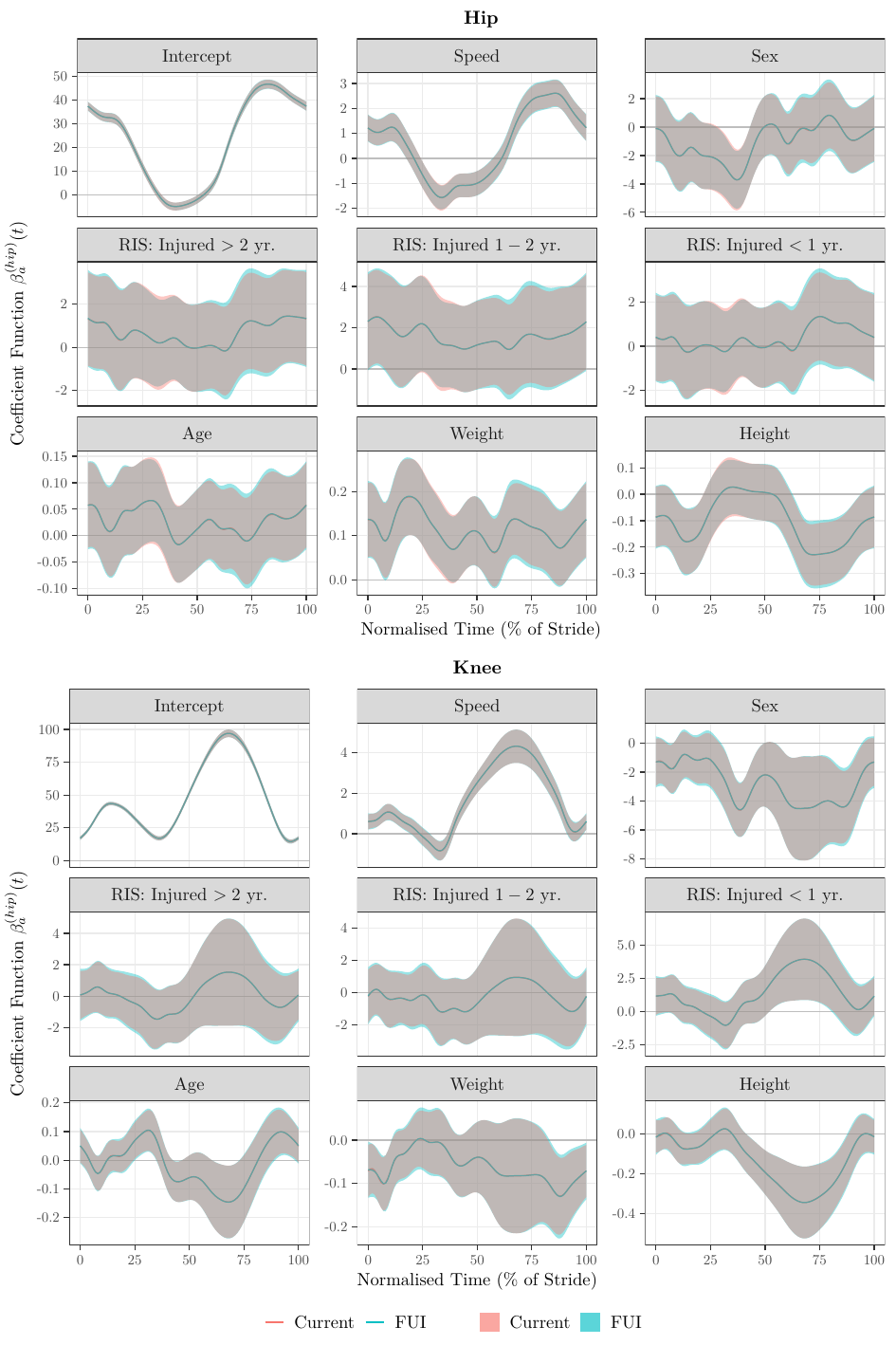}
    \caption{Comparison of fixed effects point estimates and confidence intervals between our proposed bivariate functional mixed model and separate FUI models fitted within each dimension. The confidence intervals for our current model are obtained via the Wald approach described in Section \ref{sec:bfmm-pw-conf-int}. The confidence intervals for the FUI are obtained via the analytic approach described by \textcite{cui_fast_2022}.}
    \label{fig:fui-analytic}
\end{figure}

\begin{figure}
    \centering
    \includegraphics[width = 0.8\textwidth]{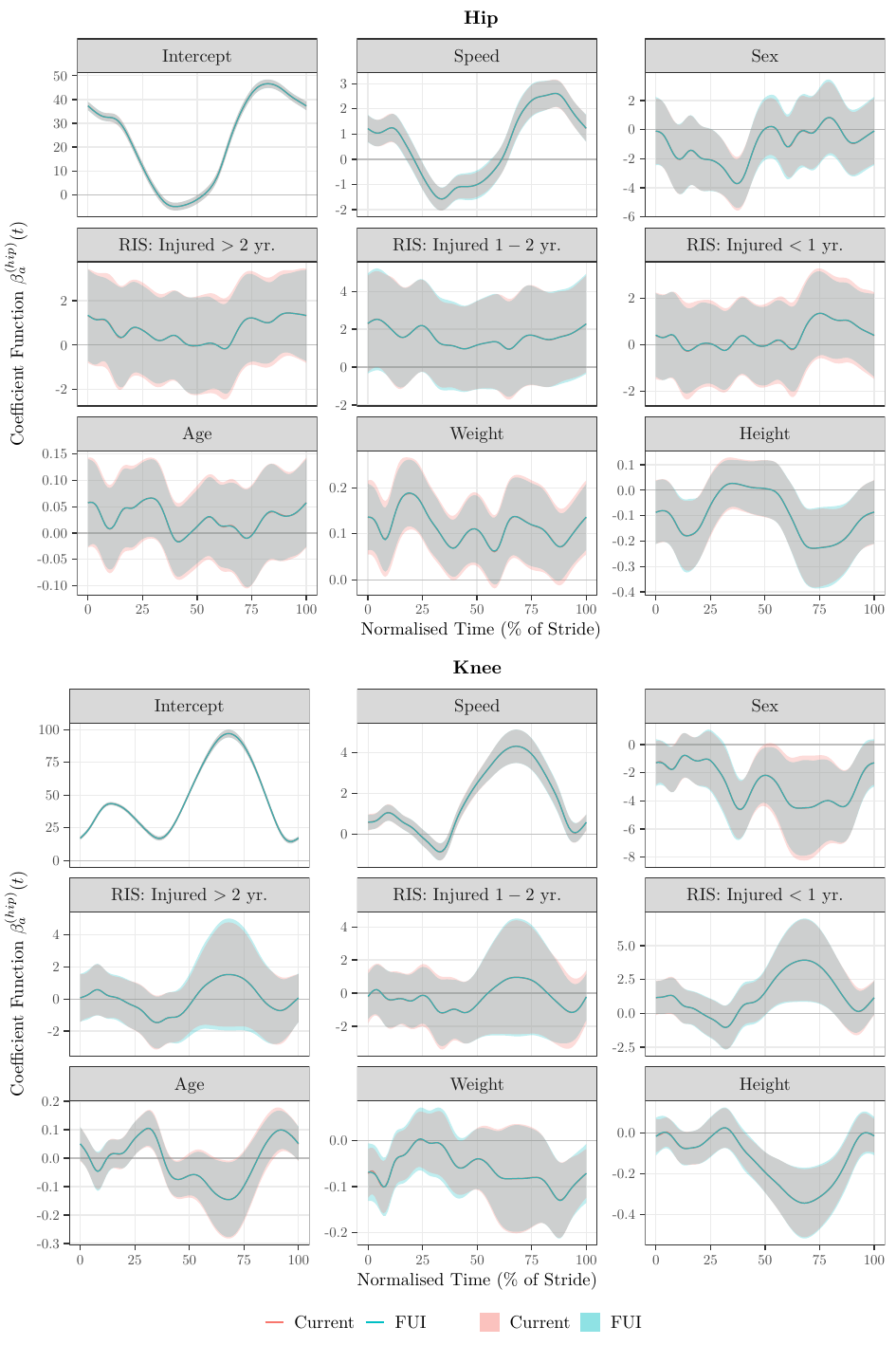}
    \caption{Comparison of fixed effects point estimates and confidence intervals between our proposed bivariate functional mixed model and separate FUI models fitted within each dimension. Confidence intervals are obtained via bootstrap approaches for the respective methods. For the comparison, $B=100$ bootstrap replicates are used for FUI, which is the default in the software implementation \parencite[Supplementary Material]{cui_fast_2022}.}
    \label{fig:fui-bootstrap}
\end{figure}

Table \ref{tab:fui-comp-time}  displays the computation times for each FUI fit.
The bootstrap approach is, as expected, more computationally intensive even after parallelising model fits.
It is substantially faster to fit separate FUI models in each dimension to obtain point estimates and pointwise confidence intervals using either approach than it is to fit multiFAMM with any settings.
On the other hand, fitting separate FUIs neglects the multivariate nature of the data and does not provide decompositions of the multivariate functional covariance structures at the different levels like multiFAMM (and to a lesser extent, our proposed model) does.
It also does not provide a way to construct joint bands for multivariate functions (though it seems possible that the bootstrap and analytic approaches could be extended for this purpose in the future).
In practice, more than $B=100$ bootstrap replicates might be also be required.
Figures \ref{fig:fui-analytic} and \ref{fig:fui-bootstrap} show comparisons of the results from our proposed model with FUI using analytic and bootstrap approaches, respectively. The point estimates, indicated as solid lines, are indistinguishable from one another. The confidence intervals, indicated by semi-transparent shaded regions with grey representing the region of overlap between the two estimates, show only very small disagreements. This does not come as a surprise, as the two approaches are philosophically similar. However, FUI post-smooths the regression coefficient functions, which, in this instance, appears have had little (if any) impact.

\clearpage


\printbibliography

@article{messier_2-year_2018,
  title = {A 2-{{Year Prospective Cohort Study}} of {{Overuse Running Injuries}}: {{The Runners}} and {{Injury Longitudinal Study}} ({{TRAILS}})},
  shorttitle = {A 2-{{Year Prospective Cohort Study}} of {{Overuse Running Injuries}}},
  author = {Messier, Stephen P. and Martin, David F. and Mihalko, Shannon L. and Ip, Edward and DeVita, Paul and Cannon, D. Wayne and Love, Monica and Beringer, Danielle and Saldana, Santiago and Fellin, Rebecca E. and Seay, Joseph F.},
  date = {2018-07},
  journaltitle = {The American Journal of Sports Medicine},
  shortjournal = {Am J Sports Med},
  volume = {46},
  number = {9},
  eprint = {29791183},
  eprinttype = {pmid},
  pages = {2211--2221},
  issn = {1552-3365},
  doi = {10.1177/0363546518773755},
  langid = {english}
}

@article{li_latent_2023,
  title = {Latent Factor Model for Multivariate Functional Data},
  author = {Li, Ruonan and Xiao, Luo},
  date = {2023},
  journaltitle = {Biometrics},
  volume = {79},
  number = {4},
  pages = {3307--3318},
  issn = {1541-0420},
  doi = {10.1111/biom.13924},
  url = {https://onlinelibrary.wiley.com/doi/abs/10.1111/biom.13924},
  urldate = {2024-08-14},
  langid = {english}
}

@article{liu_multivariate_2022,
  title = {Multivariate {{Functional Regression Via Nested Reduced-Rank Regularization}}},
  author = {Liu, Xiaokang and Ma, Shujie and Chen, Kun},
  date = {2022-01-02},
  journaltitle = {Journal of Computational and Graphical Statistics},
  volume = {31},
  number = {1},
  pages = {231--240},
  publisher = {Taylor \& Francis},
  issn = {1061-8600},
  doi = {10.1080/10618600.2021.1960850},
  url = {https://doi.org/10.1080/10618600.2021.1960850},
  urldate = {2024-08-14}
}

@article{zhu_one-way_2022,
  title = {One-Way {{MANOVA}} for Functional Data via {{Lawley}}–{{Hotelling}} Trace Test},
  author = {Zhu, Tianming and Zhang, Jin-Ting and Cheng, Ming-Yen},
  date = {2022-11-01},
  journaltitle = {Journal of Multivariate Analysis},
  shortjournal = {Journal of Multivariate Analysis},
  volume = {192},
  pages = {105095},
  issn = {0047-259X},
  doi = {10.1016/j.jmva.2022.105095},
  url = {https://www.sciencedirect.com/science/article/pii/S0047259X22000884},
  urldate = {2024-08-14}
}

@article{diquigiovanni_conformal_2022,
  title = {Conformal Prediction Bands for Multivariate Functional Data},
  author = {Diquigiovanni, Jacopo and Fontana, Matteo and Vantini, Simone},
  date = {2022-05-01},
  journaltitle = {Journal of Multivariate Analysis},
  shortjournal = {Journal of Multivariate Analysis},
  volume = {189},
  pages = {104879},
  issn = {0047-259X},
  doi = {10.1016/j.jmva.2021.104879},
  url = {https://www.sciencedirect.com/science/article/pii/S0047259X21001573},
  urldate = {2024-08-14}
}

@article{jiang_analysis_2022,
  title = {Analysis of Multivariate Non-Gaussian Functional Data: {{A}} Semiparametric Latent Process Approach},
  shorttitle = {Analysis of Multivariate Non-Gaussian Functional Data},
  author = {Jiang, Jiakun and Lin, Huazhen and Zhong, Qingzhi and Li, Yi},
  date = {2022-05-01},
  journaltitle = {Journal of Multivariate Analysis},
  shortjournal = {Journal of Multivariate Analysis},
  volume = {189},
  pages = {104888},
  issn = {0047-259X},
  doi = {10.1016/j.jmva.2021.104888},
  url = {https://www.sciencedirect.com/science/article/pii/S0047259X21001664},
  urldate = {2024-08-14}
}

@thesis{zohner_feature_2021,
  type = {phdthesis},
  title = {Feature {{Learning}} and {{Bayesian Functional Regression}} for {{High-Dimensional Complex Data}}},
  author = {Zohner, Ye Emma Mariam},
  date = {2021},
  institution = {Rice University},
  location = {United States -- Texas},
  url = {https://www.proquest.com/docview/2715479373/abstract/89C838326EDE4980PQ/1},
  urldate = {2024-05-14},
  isbn = {9798841721215},
  langid = {english},
  pagetotal = {124}
}

@article{scheipl_functional_2015,
	title = {Functional {Additive} {Mixed} {Models}},
	volume = {24},
	issn = {1061-8600},
	url = {https://www.ncbi.nlm.nih.gov/pmc/articles/PMC4560367/},
	doi = {10.1080/10618600.2014.901914},
	number = {2},
	urldate = {2021-03-09},
	journal = {Journal of Computational and Graphical Statistics},
	author = {Scheipl, Fabian and Staicu, Ana-Maria and Greven, Sonja},
	month = apr,
	year = {2015},
	pmid = {26347592},
	pmcid = {PMC4560367},
	pages = {477--501},
}

@article{zhang_testing_2017,
	title = {Testing {Gait} with {Ankle}-{Foot} {Orthoses} in {Children} with {Cerebral} {Palsy} by {Using} {Functional} {Mixed}-{Effects} {Analysis} of {Variance}},
	volume = {7},
	url = {https://www.ncbi.nlm.nih.gov/pmc/articles/PMC5594035/},
	doi = {10.1038/s41598-017-11282-1},
	language = {en},
	number = {1},
	urldate = {2020-09-16},
	journal = {Scientific Reports},
	author = {Zhang, Bairu and Twycross-Lewis, Richard and Großmann, Heiko and Morrissey, Dylan},
	year = {2017},
	pmid = {28894132},
	note = {Publisher: Nature Publishing Group},
	pages = {11081},
}

@article{warmenhoven_pca_2021,
	title = {{PCA} of waveforms and functional {PCA}: {A} primer for biomechanics},
	volume = {116},
	issn = {0021-9290},
	shorttitle = {{PCA} of waveforms and functional {PCA}},
	url = {https://www.sciencedirect.com/science/article/pii/S0021929020305303},
	doi = {10.1016/j.jbiomech.2020.110106},
	urldate = {2023-08-30},
	journal = {Journal of Biomechanics},
	author = {Warmenhoven, John and Bargary, Norma and Liebl, Dominik and Harrison, Andrew J. and Robinson, Mark A. and Gunning, Edward and Hooker, Giles},
	month = feb,
	year = {2021},
	pages = {110106},
}

@article{sergazinov_case_2023,
	title = {A case study of glucose levels during sleep using multilevel fast function on scalar regression inference},
	volume = {(Advance Online Publication https://doi.org/10.1111/biom.13878)},
	copyright = {© 2023 The International Biometric Society.},
	issn = {1541-0420},
	url = {https://onlinelibrary.wiley.com/doi/abs/10.1111/biom.13878},
	doi = {10.1111/biom.13878},
	language = {en},
	urldate = {2023-08-30},
	journal = {Biometrics},
	author = {Sergazinov, Renat and Leroux, Andrew and Cui, Erjia and Crainiceanu, Ciprian and Aurora, R. Nisha and Punjabi, Naresh M. and Gaynanova, Irina},
	month = may,
	year = {2023},
	note = {\_eprint: https://onlinelibrary.wiley.com/doi/pdf/10.1111/biom.13878},
}

@book{pinheiro_mixed-effects_2006,
	title = {Mixed-{Effects} {Models} in {S} and {S}-{PLUS}},
	isbn = {978-0-387-22747-4},
	language = {en},
	publisher = {Springer Science \& Business Media},
	author = {Pinheiro, José  S and Bates, Douglas},
	month = may,
	year = {2006},
	note = {Google-Books-ID: ZRnoBwAAQBAJ},
}

@article{greven_longitudinal_2010,
	title = {Longitudinal functional principal component analysis},
	volume = {4},
	issn = {1935-7524},
	url = {https://www.ncbi.nlm.nih.gov/pmc/articles/PMC3131008/},
	doi = {10.1214/10-EJS575},
	urldate = {2020-09-24},
	journal = {Electronic Journal of Statistics},
	author = {Greven, Sonja and Crainiceanu, Ciprian M and Caffo, Brian and Reich, Daniel},
	year = {2010},
	pmid = {21743825},
	pmcid = {PMC3131008},
	pages = {1022--1054},
}

@article{hebert-losier_one-leg_2015,
	title = {One-leg hop kinematics 20 years following anterior cruciate ligament rupture: {Data} revisited using functional data analysis},
	volume = {30},
	issn = {0268-0033},
	shorttitle = {One-leg hop kinematics 20years following anterior cruciate ligament rupture},
	url = {http://www.sciencedirect.com/science/article/pii/S0268003315002223},
	doi = {10.1016/j.clinbiomech.2015.08.010},
	language = {en},
	number = {10},
	urldate = {2020-09-16},
	journal = {Clinical Biomechanics},
	author = {Hébert-Losier, Kim and Pini, Alessia and Vantini, Simone and Strandberg, Johan and Abramowicz, Konrad and Schelin, Lina and Häger, Charlotte K.},
	month = dec,
	year = {2015},
	pages = {1153--1161},
}

@phdthesis{cederbaum_functional_2017,
	type = {Doctoral {Thesis}},
	title = {Functional {Linear} {Mixed} {Models} for {Complex} {Correlation} {Structures} and {General} {Sampling} {Grids}},
	school = {Ludwig-Maximilians-Universität München},
	author = {Cederbaum, Jona},
	month = jun,
	year = {2017},
}

@article{orendurff_little_2018,
	title = {A little bit faster: {Lower} extremity joint kinematics and kinetics as recreational runners achieve faster speeds},
	volume = {71},
	issn = {0021-9290},
	shorttitle = {A little bit faster},
	url = {https://www.sciencedirect.com/science/article/pii/S0021929018300964},
	doi = {10.1016/j.jbiomech.2018.02.010},
	language = {en},
	urldate = {2023-07-12},
	journal = {Journal of Biomechanics},
	author = {Orendurff, Michael S. and Kobayashi, Toshiki and Tulchin-Francis, Kirsten and Tullock, Ann Marie Herring and Villarosa, Chris and Chan, Charles and Kraus, Emily and Strike, Siobhan},
	month = apr,
	year = {2018},
	pages = {167--175},
}

@article{volkmann_multivariate_2021,
	title = {Multivariate functional additive mixed models},
	copyright = {© 2020 Statistical Modeling Society},
	shorttitle = {Multivariate functional additive mixed models},
	url = {https://journals.sagepub.com/doi/full/10.1177/1471082X211056158},
	doi = {10.1177/1471082X211056158},
	language = {en},
	urldate = {2021-12-09},
	journal = {Statistical Modelling},
	author = {Volkmann, Alexander and Stöcker, Almond and Scheipl, Fabian and Greven, Sonja},
	month = dec,
	year = {2021},
	note = {Publisher: SAGE PublicationsSage India: New Delhi, India},
}

@article{matabuena_estimating_2023,
	title = {Estimating {Knee} {Movement} {Patterns} of {Recreational} {Runners} {Across} {Training} {Sessions} {Using} {Multilevel} {Functional} {Regression} {Models}},
	volume = {77},
	issn = {0003-1305},
	url = {https://doi.org/10.1080/00031305.2022.2105950},
	doi = {10.1080/00031305.2022.2105950},
	number = {2},
	urldate = {2023-07-11},
	journal = {The American Statistician},
	author = {Matabuena, Marcos and Karas, Marta and Riazati, Sherveen and Caplan, Nick and Hayes, Philip R.},
	month = apr,
	year = {2023},
	note = {Publisher: Taylor \& Francis
\_eprint: https://doi.org/10.1080/00031305.2022.2105950},
	pages = {169--181},
}

@misc{ramsay_fda_2020,
	title = {{fda}: {Functional} {Data} {Analysis}. {R} package version 5.5.1. {https://CRAN.R-project.org/package=fda}},
	copyright = {GPL-2 {\textbar} GPL-3 [expanded from: GPL (≥ 2)]},
	shorttitle = {fda},
	url = {https://CRAN.R-project.org/package=fda},
	urldate = {2020-10-10},
	author = {Ramsay, James O. and Graves, Spencer and Hooker, Giles},
	month = aug,
	year = {2020},
}

@misc{golovkine_use_2023,
	title = {On the use of the {Gram} matrix for multivariate functional principal components analysis {[arXiv:2306.12949 [stat]]}},
	url = {http://arxiv.org/abs/2306.12949},
	doi = {10.48550/arXiv.2306.12949},
	urldate = {2023-06-27},
	publisher = {arXiv},
	author = {Golovkine, Steven and Gunning, Edward and Simpkin, Andrew J. and Bargary, Norma},
	month = jun,
	year = {2023},
	note = {arXiv:2306.12949 [stat]},
}

@article{crainiceanu_bootstrap-based_2012,
	title = {Bootstrap-based inference on the difference in the means of two correlated functional processes},
	volume = {31},
	issn = {0277-6715},
	url = {https://www.ncbi.nlm.nih.gov/pmc/articles/PMC3966027/},
	doi = {10.1002/sim.5439},
	number = {26},
	urldate = {2022-01-10},
	journal = {Statistics in Medicine},
	author = {Crainiceanu, Ciprian M. and Staicu, Ana-Maria and Ray, Shubankar and Punjabi, Naresh},
	month = nov,
	year = {2012},
	pmid = {22855258},
	pmcid = {PMC3966027},
	pages = {3223--3240},
}

@misc{pinheiro_nlme_2022,
	title = {{nlme}: {Linear} and {Nonlinear} {Mixed} {Effects} {Models}. {R} package version 3.1-155. {https://CRAN.R-project.org/package=nlme}},
	copyright = {GPL-2 {\textbar} GPL-3 [expanded from: GPL (≥ 2)]},
	shorttitle = {nlme},
	url = {https://CRAN.R-project.org/package=nlme},
	urldate = {2022-06-03},
	author = {Pinheiro, José  S and Bates, Douglas and DebRoy, Saikat and Sarkar, Deepayan and Heisterkamp, Siem and Van Willigen, Bert and Ranke, Johannes},
	month = mar,
	year = {2022},
}

@article{wood_fast_2011,
	title = {Fast stable restricted maximum likelihood and marginal likelihood estimation of semiparametric generalized linear models},
	volume = {73},
	copyright = {© 2010 Royal Statistical Society},
	issn = {1467-9868},
	url = {https://onlinelibrary.wiley.com/doi/abs/10.1111/j.1467-9868.2010.00749.x},
	doi = {10.1111/j.1467-9868.2010.00749.x},
	language = {en},
	number = {1},
	urldate = {2023-06-05},
	journal = {Journal of the Royal Statistical Society Series B: Statistical Methodology},
	author = {Wood, Simon N.},
	year = {2011},
	note = {\_eprint: https://onlinelibrary.wiley.com/doi/pdf/10.1111/j.1467-9868.2010.00749.x},
	pages = {3--36},
}

@article{liu_functional_2012,
	title = {Functional mixed effects models},
	volume = {4},
	copyright = {Copyright © 2012 Wiley Periodicals, Inc.},
	issn = {1939-0068},
	url = {https://onlinelibrary.wiley.com/doi/abs/10.1002/wics.1226},
	doi = {10.1002/wics.1226},
	language = {en},
	number = {6},
	urldate = {2021-03-09},
	journal = {WIREs Computational Statistics},
	author = {Liu, Ziyue and Guo, Wensheng},
	year = {2012},
	note = {\_eprint: https://onlinelibrary.wiley.com/doi/pdf/10.1002/wics.1226},
	pages = {527--534},
}

@article{degras_simultaneous_2017,
	title = {Simultaneous confidence bands for the mean of functional data},
	volume = {9},
	copyright = {© 2017 The Authors. WIREs Computational Statistics published by Wiley Periodicals, Inc.},
	issn = {1939-0068},
	url = {https://onlinelibrary.wiley.com/doi/abs/10.1002/wics.1397},
	doi = {10.1002/wics.1397},
	language = {en},
	number = {3},
	urldate = {2021-03-24},
	journal = {WIREs Computational Statistics},
	author = {Degras, David},
	year = {2017},
	pages = {e1397},
}

@article{cui_fast_2023,
	title = {Fast {Multilevel} {Functional} {Principal} {Component} {Analysis}},
	volume = {32},
	issn = {1061-8600},
	url = {https://doi.org/10.1080/10618600.2022.2115500},
	doi = {10.1080/10618600.2022.2115500},
	number = {2},
	urldate = {2023-06-05},
	journal = {Journal of Computational and Graphical Statistics},
	author = {Cui, Erjia and Li, Ruonan and Crainiceanu, Ciprian M. and Xiao, Luo},
	month = apr,
	year = {2023},
	note = {Publisher: Taylor \& Francis
\_eprint: https://doi.org/10.1080/10618600.2022.2115500},
	pages = {366--377},
}

@article{park_longitudinal_2015,
	title = {Longitudinal functional data analysis},
	volume = {4},
	issn = {2049-1573},
	url = {https://onlinelibrary.wiley.com/doi/abs/10.1002/sta4.89},
	doi = {10.1002/sta4.89},
	language = {en},
	number = {1},
	urldate = {2023-03-16},
	journal = {Stat},
	author = {Park, So Young and Staicu, Ana-Maria},
	year = {2015},
	note = {\_eprint: https://onlinelibrary.wiley.com/doi/pdf/10.1002/sta4.89},
	pages = {212--226},
}

@article{li_fast_2020,
	title = {Fast covariance estimation for multivariate sparse functional data},
	volume = {9},
	issn = {2049-1573},
	url = {https://onlinelibrary.wiley.com/doi/abs/10.1002/sta4.245},
	doi = {10.1002/sta4.245},
	language = {en},
	number = {1},
	urldate = {2023-03-15},
	journal = {Stat},
	author = {Li, Cai and Xiao, Luo and Luo, Sheng},
	year = {2020},
	note = {\_eprint: https://onlinelibrary.wiley.com/doi/pdf/10.1002/sta4.245},
	pages = {e245},
}

@misc{genz_mvtnorm_2021,
	title = {{mvtnorm}: {Multivariate} {Normal} and t {Distributions}. {R} package version 1.1-3. {http://CR} {AN.R-project.org/package=mvtnorm}},
	copyright = {GPL-2},
	shorttitle = {mvtnorm},
	url = {https://CRAN.R-project.org/package=mvtnorm},
	urldate = {2023-02-16},
	author = {Genz, Alan and Bretz, Frank and Miwa, Tetsuhisa and Mi, Xuefei and Leisch, Friedrich and Scheipl, Fabian and Hothorn, Torsten},
	month = oct,
	year = {2021},
}

@misc{r_core_team_r_2022,
	address = {Vienna, Austria},
	title = {R: {A} {Language} and {Environment} for {Statistical} {Computing}},
	url = {https://www.R-project.org/},
	publisher = {R Foundation for Statistical Computing},
	author = {{R Core Team}},
	year = {2022},
}

@article{di_multilevel_2009,
	title = {Multilevel functional principal component analysis},
	volume = {3},
	issn = {1932-6157},
	url = {https://www.ncbi.nlm.nih.gov/pmc/articles/PMC2835171/},
	doi = {10.1214/08-AOAS206SUPP},
	number = {1},
	urldate = {2020-08-23},
	journal = {The Annals of Applied Statistics},
	author = {Di, Chong-Zhi and Crainiceanu, Ciprian M. and Caffo, Brian S. and Punjabi, Naresh M.},
	month = mar,
	year = {2009},
	pmid = {20221415},
	pmcid = {PMC2835171},
	pages = {458--488},
}

@article{zhu_multivariate_2017,
	title = {Multivariate functional response regression, with application to fluorescence spectroscopy in a cervical pre-cancer study},
	volume = {111},
	issn = {0167-9473},
	url = {https://www.sciencedirect.com/science/article/pii/S0167947317300245},
	doi = {10.1016/j.csda.2017.02.004},
	language = {en},
	urldate = {2023-01-20},
	journal = {Computational Statistics \& Data Analysis},
	author = {Zhu, Hongxiao and Morris, Jeffrey S. and Wei, Fengrong and Cox, Dennis D.},
	month = jul,
	year = {2017},
	pages = {88--101},
}

@article{taylor_repeatability_2010,
	title = {Repeatability and reproducibility of {OSSCA}, a functional approach for assessing the kinematics of the lower limb},
	volume = {32},
	issn = {0966-6362},
	url = {https://www.sciencedirect.com/science/article/pii/S0966636210001293},
	doi = {10.1016/j.gaitpost.2010.05.005},
	language = {en},
	number = {2},
	urldate = {2023-01-17},
	journal = {Gait \& Posture},
	author = {Taylor, W. R. and Kornaropoulos, E. I. and Duda, G. N. and Kratzenstein, S. and Ehrig, R. M. and Arampatzis, A. and Heller, M. O.},
	month = jun,
	year = {2010},
	pages = {231--236},
}

@book{grimshaw_bios_2007,
	title = {{BIOS} {Instant} {Notes} in {Sport} and {Exercise} {Biomechanics}},
	isbn = {978-0-203-48830-0},
	url = {https://www.taylorfrancis.com/books/mono/10.4324/9780203488300/bios-instant-notes-sport-exercise-biomechanics-paul-grimshaw-neil-fowler-adrian-lees-adrian-burden},
	language = {en},
	urldate = {2023-01-14},
	publisher = {Routledge},
	author = {Grimshaw, Paul and Fowler, Neil and Lees, Adrian and Burden, Adrian},
	month = apr,
	year = {2007},
	doi = {10.4324/9780203488300},
}

@article{hespanhol_junior_health_2017,
	title = {Health and {Economic} {Burden} of {Running}-{Related} {Injuries} in {Dutch} {Trailrunners}: {A} {Prospective} {Cohort} {Study}},
	volume = {47},
	issn = {1179-2035},
	shorttitle = {Health and {Economic} {Burden} of {Running}-{Related} {Injuries} in {Dutch} {Trailrunners}},
	doi = {10.1007/s40279-016-0551-8},
	language = {eng},
	number = {2},
	journal = {Sports Medicine (Auckland, N.Z.)},
	author = {Hespanhol Junior, Luiz Carlos and van Mechelen, Willem and Verhagen, Evert},
	month = feb,
	year = {2017},
	pmid = {27222128},
	pmcid = {PMC5266769},
	pages = {367--377},
}

@article{ceyssens_biomechanical_2019,
	title = {Biomechanical {Risk} {Factors} {Associated} with {Running}-{Related} {Injuries}: {A} {Systematic} {Review}},
	volume = {49},
	issn = {1179-2035},
	shorttitle = {Biomechanical {Risk} {Factors} {Associated} with {Running}-{Related} {Injuries}},
	url = {https://doi.org/10.1007/s40279-019-01110-z},
	doi = {10.1007/s40279-019-01110-z},
	language = {en},
	number = {7},
	urldate = {2022-11-18},
	journal = {Sports Medicine},
	author = {Ceyssens, Linde and Vanelderen, Romy and Barton, Christian and Malliaras, Peter and Dingenen, Bart},
	month = jul,
	year = {2019},
	pages = {1095--1115},
}

@article{willwacher_running-related_2022,
	title = {Running-{Related} {Biomechanical} {Risk} {Factors} for {Overuse} {Injuries} in {Distance} {Runners}: {A} {Systematic} {Review} {Considering} {Injury} {Specificity} and the {Potentials} for {Future} {Research}},
	volume = {52},
	issn = {1179-2035},
	shorttitle = {Running-{Related} {Biomechanical} {Risk} {Factors} for {Overuse} {Injuries} in {Distance} {Runners}},
	url = {https://doi.org/10.1007/s40279-022-01666-3},
	doi = {10.1007/s40279-022-01666-3},
	language = {en},
	number = {8},
	urldate = {2022-11-18},
	journal = {Sports Medicine},
	author = {Willwacher, Steffen and Kurz, Markus and Robbin, Johanna and Thelen, Matthias and Hamill, Joseph and Kelly, Luke and Mai, Patrick},
	month = aug,
	year = {2022},
	pages = {1863--1877},
}

@book{fieller_basics_2016,
	address = {New York},
	title = {Basics of {Matrix} {Algebra} for {Statistics} with {R}},
	isbn = {978-1-315-37020-0},
	publisher = {Chapman {\&} Hall/CRC},
	author = {Fieller, Nick},
	month = dec,
	year = {2016},
	doi = {10.1201/9781315370200},
}

@article{park_simple_2018,
	title = {Simple fixed-effects inference for complex functional models},
	volume = {19},
	issn = {1465-4644},
	url = {https://doi.org/10.1093/biostatistics/kxx026},
	doi = {10.1093/biostatistics/kxx026},
	number = {2},
	urldate = {2022-10-22},
	journal = {Biostatistics},
	author = {Park, So Young and Staicu, Ana-Maria and Xiao, Luo and Crainiceanu, Ciprian M},
	month = apr,
	year = {2018},
	pages = {137--152},
}

@article{becker_biomechanical_2017,
	title = {Biomechanical {Factors} {Associated} {With} {Achilles} {Tendinopathy} and {Medial} {Tibial} {Stress} {Syndrome} in {Runners}},
	volume = {45},
	issn = {0363-5465},
	url = {https://doi.org/10.1177/0363546517708193},
	doi = {10.1177/0363546517708193},
	language = {en},
	number = {11},
	urldate = {2022-10-20},
	journal = {The American Journal of Sports Medicine},
	author = {Becker, James and James, Stanley and Wayner, Robert and Osternig, Louis and Chou, Li-Shan},
	month = sep,
	year = {2017},
	note = {Publisher: SAGE Publications Inc STM},
	pages = {2614--2621},
}

@article{bramah_is_2018,
	title = {Is {There} a {Pathological} {Gait} {Associated} {With} {Common} {Soft} {Tissue} {Running} {Injuries}?},
	volume = {46},
	issn = {0363-5465},
	url = {https://doi.org/10.1177/0363546518793657},
	doi = {10.1177/0363546518793657},
	language = {en},
	number = {12},
	urldate = {2022-10-20},
	journal = {The American Journal of Sports Medicine},
	author = {Bramah, Christopher and Preece, Stephen J. and Gill, Niamh and Herrington, Lee},
	month = oct,
	year = {2018},
	note = {Publisher: SAGE Publications Inc STM},
	pages = {3023--3031},
}

@article{mann_association_2015,
	title = {Association of previous injury and speed with running style and stride-to-stride fluctuations},
	volume = {25},
	issn = {1600-0838},
	url = {https://onlinelibrary.wiley.com/doi/abs/10.1111/sms.12397},
	doi = {10.1111/sms.12397},
	language = {en},
	number = {6},
	urldate = {2022-10-20},
	journal = {Scandinavian Journal of Medicine \& Science in Sports},
	author = {Mann, R. and Malisoux, L. and Nührenbörger, C. and Urhausen, A. and Meijer, K. and Theisen, D.},
	year = {2015},
	note = {\_eprint: https://onlinelibrary.wiley.com/doi/pdf/10.1111/sms.12397},
	pages = {e638--e645},
}

@article{kenward_small_1997,
	title = {Small {Sample} {Inference} for {Fixed} {Effects} from {Restricted} {Maximum} {Likelihood}},
	volume = {53},
	issn = {0006-341X},
	url = {https://www.jstor.org/stable/2533558},
	doi = {10.2307/2533558},
	number = {3},
	urldate = {2022-10-19},
	journal = {Biometrics},
	author = {Kenward, Michael G. and Roger, James H.},
	year = {1997},
	note = {Publisher: [Wiley, International Biometric Society]},
	pages = {983--997},
}

@article{burke_comparison_2022,
	title = {Comparison of impact accelerations between injury-resistant and recently injured recreational runners},
	volume = {17},
	issn = {1932-6203},
	url = {https://journals.plos.org/plosone/article?id=10.1371/journal.pone.0273716},
	doi = {10.1371/journal.pone.0273716},
	language = {en},
	number = {9},
	urldate = {2022-10-13},
	journal = {PLOS ONE},
	author = {Burke, Aoife and Dillon, Sarah and O’Connor, Siobhán and Whyte, Enda F. and Gore, Shane and Moran, Kieran A.},
	month = sep,
	year = {2022},
	note = {Publisher: Public Library of Science},
	pages = {e0273716},
}

@article{saragiotto_what_2014,
	title = {What are the {Main} {Risk} {Factors} for {Running}-{Related} {Injuries}?},
	volume = {44},
	issn = {1179-2035},
	url = {https://doi.org/10.1007/s40279-014-0194-6},
	doi = {10.1007/s40279-014-0194-6},
	language = {en},
	number = {8},
	urldate = {2022-10-13},
	journal = {Sports Medicine},
	author = {Saragiotto, Bruno Tirotti and Yamato, Tiê Parma and Hespanhol Junior, Luiz Carlos and Rainbow, Michael J. and Davis, Irene S. and Lopes, Alexandre Dias},
	month = aug,
	year = {2014},
	pages = {1153--1163},
}

@article{aguilera-morillo_multi-class_2020,
	title = {Multi-class classification of biomechanical data: {A} functional {LDA} approach based on multi-class penalized functional {PLS}},
	volume = {20},
	issn = {1471-082X},
	shorttitle = {Multi-class classification of biomechanical data},
	url = {https://doi.org/10.1177/1471082X19871157},
	doi = {10.1177/1471082X19871157},
	language = {en},
	number = {6},
	urldate = {2022-10-03},
	journal = {Statistical Modelling},
	author = {Aguilera-Morillo, M. Carmen and Aguilera, Ana M.},
	month = dec,
	year = {2020},
	note = {Publisher: SAGE Publications India},
	pages = {592--616},
}

@article{glazier_beyond_2021,
	title = {Beyond animated skeletons: {How} can biomechanical feedback be used to enhance sports performance?},
	volume = {129},
	issn = {0021-9290},
	shorttitle = {Beyond animated skeletons},
	url = {https://www.sciencedirect.com/science/article/pii/S0021929021004553},
	doi = {10.1016/j.jbiomech.2021.110686},
	language = {en},
	urldate = {2021-10-13},
	journal = {Journal of Biomechanics},
	author = {Glazier, Paul S.},
	month = dec,
	year = {2021},
	pages = {110686},
}

@article{dillon_injury-resistant_2021,
	title = {Do {Injury}-{Resistant} {Runners} {Have} {Distinct} {Differences} in {Clinical} {Measures} {Compared} with {Recently} {Injured} {Runners}?},
	volume = {53},
	issn = {1530-0315},
	doi = {10.1249/MSS.0000000000002649},
	language = {eng},
	number = {9},
	journal = {Medicine and Science in Sports and Exercise},
	author = {Dillon, Sarah and Burke, Aoife and Whyte, Enda F. and O'Connor, Siobhán and Gore, Shane and Moran, Kieran A.},
	month = sep,
	year = {2021},
	pmid = {33899779},
	pages = {1807--1817},
}

@incollection{lamb_assessing_2017,
	edition = {2},
	title = {Assessing movement coordination},
	isbn = {978-0-203-09554-6},
	booktitle = {Biomechanical {Evaluation} of {Movement} in {Sport} and {Exercise}},
	publisher = {Routledge},
	author = {Lamb, Peter F. and Bartlett, Roger M.},
	year = {2017},
	note = {Num Pages: 22},
}

@article{friedman_sparse_2008,
	title = {Sparse inverse covariance estimation with the graphical lasso},
	volume = {9},
	issn = {1465-4644},
	url = {https://doi.org/10.1093/biostatistics/kxm045},
	doi = {10.1093/biostatistics/kxm045},
	number = {3},
	urldate = {2022-07-26},
	journal = {Biostatistics},
	author = {Friedman, Jerome and Hastie, Trevor and Tibshirani, Robert},
	month = jul,
	year = {2008},
	pages = {432--441},
}

@article{fieuws_pairwise_2006,
	title = {Pairwise {Fitting} of {Mixed} {Models} for the {Joint} {Modeling} of {Multivariate} {Longitudinal} {Profiles}},
	volume = {62},
	issn = {1541-0420},
	url = {https://onlinelibrary.wiley.com/doi/abs/10.1111/j.1541-0420.2006.00507.x},
	doi = {10.1111/j.1541-0420.2006.00507.x},
	language = {en},
	number = {2},
	urldate = {2022-07-26},
	journal = {Biometrics},
	author = {Fieuws, Steffen and Verbeke, Geert},
	year = {2006},
	note = {\_eprint: https://onlinelibrary.wiley.com/doi/pdf/10.1111/j.1541-0420.2006.00507.x},
	pages = {424--431},
}

@article{morris_using_2019,
	title = {Using simulation studies to evaluate statistical methods},
	volume = {38},
	issn = {1097-0258},
	url = {https://onlinelibrary.wiley.com/doi/abs/10.1002/sim.8086},
	doi = {10.1002/sim.8086},
	language = {en},
	number = {11},
	urldate = {2022-07-07},
	journal = {Statistics in Medicine},
	author = {Morris, Tim P. and White, Ian R. and Crowther, Michael J.},
	year = {2019},
	note = {\_eprint: https://onlinelibrary.wiley.com/doi/pdf/10.1002/sim.8086},
	pages = {2074--2102},
}

@article{cui_fast_2022,
	title = {Fast {Univariate} {Inference} for {Longitudinal} {Functional} {Models}},
	volume = {31},
	issn = {1061-8600},
	url = {https://doi.org/10.1080/10618600.2021.1950006},
	doi = {10.1080/10618600.2021.1950006},
	number = {1},
	urldate = {2022-05-31},
	journal = {Journal of Computational and Graphical Statistics},
	author = {Cui, Erjia and Leroux, Andrew and Smirnova, Ekaterina and Crainiceanu, Ciprian M.},
	month = jan,
	year = {2022},
	note = {Publisher: Taylor \& Francis
\_eprint: https://doi.org/10.1080/10618600.2021.1950006},
	pages = {219--230},
}

@article{hadjipantelis_unifying_2015,
	title = {Unifying {Amplitude} and {Phase} {Analysis}: {A} {Compositional} {Data} {Approach} to {Functional} {Multivariate} {Mixed}-{Effects} {Modeling} of {Mandarin} {Chinese}},
	volume = {110},
	issn = {0162-1459},
	shorttitle = {Unifying {Amplitude} and {Phase} {Analysis}},
	url = {https://doi.org/10.1080/01621459.2015.1006729},
	doi = {10.1080/01621459.2015.1006729},
	number = {510},
	urldate = {2022-05-08},
	journal = {Journal of the American Statistical Association},
	author = {Hadjipantelis, P. Z. and Aston, J. A. D. and Müller, H. G. and Evans, J. P.},
	month = apr,
	year = {2015},
	pmid = {26692591},
	note = {Publisher: Taylor \& Francis
\_eprint: https://doi.org/10.1080/01621459.2015.1006729},
	pages = {545--559},
}

@article{lee_bayesian_2019,
	title = {Bayesian {Semiparametric} {Functional} {Mixed} {Models} for {Serially} {Correlated} {Functional} {Data}, {With} {Application} to {Glaucoma} {Data}},
	volume = {114},
	issn = {0162-1459},
	url = {https://doi.org/10.1080/01621459.2018.1476242},
	doi = {10.1080/01621459.2018.1476242},
	number = {526},
	urldate = {2022-04-28},
	journal = {Journal of the American Statistical Association},
	author = {Lee, Wonyul and Miranda, Michelle F. and Rausch, Philip and Baladandayuthapani, Veerabhadran and Fazio, Massimo and Downs, J. Crawford and Morris, Jeffrey S.},
	month = apr,
	year = {2019},
	pmid = {31235987},
	note = {Publisher: Taylor \& Francis
\_eprint: https://doi.org/10.1080/01621459.2018.1476242},
	pages = {495--513},
}

@article{pataky_zero-_2015,
	title = {Zero- vs. one-dimensional, parametric vs. non-parametric, and confidence interval vs. hypothesis testing procedures in one-dimensional biomechanical trajectory analysis},
	volume = {48},
	issn = {0021-9290},
	url = {https://www.sciencedirect.com/science/article/pii/S0021929015001438},
	doi = {10.1016/j.jbiomech.2015.02.051},
	language = {en},
	number = {7},
	urldate = {2022-03-07},
	journal = {Journal of Biomechanics},
	author = {Pataky, Todd C. and Vanrenterghem, Jos and Robinson, Mark A.},
	month = may,
	year = {2015},
	pages = {1277--1285},
}

@article{aston_linguistic_2010,
	title = {Linguistic {Pitch} {Analysis} using {Functional} {Principal} {Component} {Mixed} {Effect} {Models}},
	volume = {59},
	issn = {0035-9254},
	url = {https://www.jstor.org/stable/40541687},
	number = {2},
	urldate = {2022-03-01},
	journal = {Journal of the Royal Statistical Society Series C: Applied Statistics},
	author = {Aston, John A. D. and Chiou, Jeng-Min and Evans, Jonathan P.},
	year = {2010},
	note = {Publisher: [Wiley, Royal Statistical Society]},
	pages = {297--317},
}

@book{winter_biomechanics_1979,
	title = {Biomechanics of {Human} {Movement}},
	isbn = {978-0-471-03476-6},
	language = {en},
	publisher = {Wiley},
	author = {Winter, David A.},
	year = {1979},
}

@misc{volkmann_multifamm_2021,
	title = {{multifamm}: {Multivariate} {Functional} {Additive} {Mixed} {Models}. {R} package version 0.1.1. {https://CRAN.R-project.org/package=multi} {famm}},
	copyright = {GPL-2 {\textbar} GPL-3 [expanded from: GPL (≥ 2)]},
	shorttitle = {multifamm},
	url = {https://CRAN.R-project.org/package=multifamm},
	urldate = {2022-02-08},
	author = {Volkmann, Alexander},
	month = sep,
	year = {2021},
}

@article{faraway_regression_1997,
	title = {Regression {Analysis} for a {Functional} {Response}},
	volume = {39},
	issn = {0040-1706},
	url = {https://www.jstor.org/stable/1271130},
	doi = {10.2307/1271130},
	number = {3},
	urldate = {2022-01-11},
	journal = {Technometrics},
	author = {Faraway, Julian J.},
	year = {1997},
	note = {Publisher: [Taylor \& Francis, Ltd., American Statistical Association, American Society for Quality]},
	pages = {254--261},
}

@book{wood_generalized_2017,
	address = {Boca Raton},
	edition = {2},
	title = {Generalized {Additive} {Models}: {An} {Introduction} with {R}},
	isbn = {978-1-315-37027-9},
	shorttitle = {Generalized {Additive} {Models}},
	publisher = {Chapman {\&} Hall/CRC},
	author = {Wood, Simon N.},
	month = may,
	year = {2017},
	doi = {10.1201/9781315370279},
}

@article{fan_two-step_2000,
	title = {Two-{Step} {Estimation} of {Functional} {Linear} {Models} with {Applications} to {Longitudinal} {Data}},
	volume = {62},
	issn = {1369-7412},
	url = {https://www.jstor.org/stable/3088861},
	number = {2},
	urldate = {2022-01-08},
	journal = {Journal of the Royal Statistical Society Series B: Statistical Methodology},
	author = {Fan, Jianqing and Zhang, Jin-Ting},
	year = {2000},
	note = {Publisher: [Royal Statistical Society, Wiley]},
	pages = {303--322},
}

@article{bauer_introduction_2018,
	title = {An introduction to semiparametric function-on-scalar regression},
	volume = {18},
	issn = {1471-082X},
	url = {https://doi.org/10.1177/1471082X17748034},
	doi = {10.1177/1471082X17748034},
	language = {en},
	number = {3-4},
	urldate = {2021-11-25},
	journal = {Statistical Modelling},
	author = {Bauer, Alexander and Scheipl, Fabian and Küchenhoff, Helmut and Gabriel, Alice-Agnes},
	month = jun,
	year = {2018},
	note = {Publisher: SAGE Publications India},
	pages = {346--364},
}

@article{morris_comparison_2017,
	title = {Comparison and contrast of two general functional regression modelling frameworks},
	volume = {17},
	issn = {1471-082X},
	url = {https://doi.org/10.1177/1471082X16681875},
	doi = {10.1177/1471082X16681875},
	language = {en},
	number = {1-2},
	urldate = {2021-11-25},
	journal = {Statistical Modelling},
	author = {Morris, Jeffrey S.},
	month = feb,
	year = {2017},
	note = {Publisher: SAGE Publications India},
	pages = {59--85},
}

@article{morris_wavelet-based_2006,
	title = {Wavelet-based functional mixed models},
	volume = {68},
	issn = {1369-7412},
	doi = {10.1111/j.1467-9868.2006.00539.x},
	language = {eng},
	number = {2},
	journal = {Journal of the Royal Statistical Society Series B: Statistical Methodology},
	author = {Morris, Jeffrey S. and Carroll, Raymond J.},
	month = apr,
	year = {2006},
	pmid = {19759841},
	pmcid = {PMC2744105},
	pages = {179--199},
}

@article{happ-kurz_object-oriented_2020,
	title = {Object-{Oriented} {Software} for {Functional} {Data}},
	volume = {93},
	copyright = {Copyright (c) 2020 Clara Happ-Kurz},
	issn = {1548-7660},
	url = {https://www.jstatsoft.org/index.php/jss/article/view/v093i05},
	doi = {10.18637/jss.v093.i05},
	language = {en},
	number = {1},
	urldate = {2021-05-18},
	journal = {Journal of Statistical Software},
	author = {Happ-Kurz, Clara},
	month = apr,
	year = {2020},
	note = {Number: 1},
	pages = {1--38},
}

@article{laird_random-effects_1982,
	title = {Random-{Effects} {Models} for {Longitudinal} {Data}},
	volume = {38},
	issn = {0006-341X},
	url = {https://www.jstor.org/stable/2529876},
	doi = {10.2307/2529876},
	number = {4},
	urldate = {2021-05-14},
	journal = {Biometrics},
	author = {Laird, Nan M. and Ware, James H.},
	year = {1982},
	note = {Publisher: [Wiley, International Biometric Society]},
	pages = {963--974},
}

@book{ramsay_functional_2005,
	address = {New York},
	edition = {2},
	series = {Springer {Series} in {Statistics}},
	title = {Functional {Data} {Analysis}},
	isbn = {978-0-387-40080-8},
	url = {https://www.springer.com/gp/book/9780387400808},
	language = {en},
	urldate = {2020-08-12},
	publisher = {Springer-Verlag},
	author = {Ramsay, James O. and Silverman, B. W.},
	year = {2005},
	doi = {10.1007/b98888},
}

@article{guo_functional_2002,
	title = {Functional mixed effects models},
	volume = {58},
	issn = {0006-341X},
	doi = {10.1111/j.0006-341x.2002.00121.x},
	language = {eng},
	number = {1},
	journal = {Biometrics},
	author = {Guo, Wensheng},
	month = mar,
	year = {2002},
	pmid = {11890306},
	pages = {121--128},
}

@article{trounson_effects_2020,
	title = {Effects of acute wearable resistance loading on overground running lower body kinematics},
	volume = {15},
	issn = {1932-6203},
	url = {https://journals.plos.org/plosone/article?id=10.1371/journal.pone.0244361},
	doi = {10.1371/journal.pone.0244361},
	language = {en},
	number = {12},
	urldate = {2021-03-16},
	journal = {PLOS ONE},
	author = {Trounson, Karl M. and Busch, Aglaja and Collier, Neil French and Robertson, Sam},
	month = dec,
	year = {2020},
	note = {Publisher: Public Library of Science},
	pages = {e0244361},
}

@article{goldsmith_assessing_2016,
	title = {Assessing systematic effects of stroke on motorcontrol by using hierarchical function-on-scalar regression},
	volume = {65},
	issn = {0035-9254},
	doi = {10.1111/rssc.12115},
	language = {eng},
	number = {2},
	journal = {Journal of the Royal Statistical Society Series C: Applied Statistics},
	author = {Goldsmith, Jeff and Kitago, Tomoko},
	month = feb,
	year = {2016},
	pmid = {27546913},
	pmcid = {PMC4988692},
	pages = {215--236},
}

@article{morris_functional_2015,
	title = {Functional {Regression}},
	volume = {2},
	journal = {Annual Review of Statistics and Its Application},
	author = {Morris, Jeffrey S.},
	month = mar,
	year = {2015},
	pages = {321--359},
}

@article{ferber_gait_2016,
	title = {Gait biomechanics in the era of data science},
	volume = {49},
	issn = {1873-2380},
	doi = {10.1016/j.jbiomech.2016.10.033},
	language = {eng},
	number = {16},
	journal = {Journal of Biomechanics},
	author = {Ferber, Reed and Osis, Sean T. and Hicks, Jennifer L. and Delp, Scott L.},
	month = dec,
	year = {2016},
	pmid = {27814971},
	pmcid = {PMC5407492},
	pages = {3759--3761},
}

@article{helwig_methods_2011,
	title = {Methods to temporally align gait cycle data},
	volume = {44},
	issn = {1873-2380},
	doi = {10.1016/j.jbiomech.2010.09.015},
	language = {eng},
	number = {3},
	journal = {Journal of Biomechanics},
	author = {Helwig, Nathaniel E. and Hong, Sungjin and Hsiao-Wecksler, Elizabeth T. and Polk, John D.},
	month = feb,
	year = {2011},
	pmid = {20887992},
	pages = {561--566},
}

@article{kneip_statistical_1992,
	title = {Statistical {Tools} to {Analyze} {Data} {Representing} a {Sample} of {Curves}},
	volume = {20},
	issn = {0090-5364},
	url = {https://www.jstor.org/stable/2242012},
	number = {3},
	urldate = {2021-02-21},
	journal = {The Annals of Statistics},
	author = {Kneip, Alois and Gasser, Theo},
	year = {1992},
	note = {Publisher: Institute of Mathematical Statistics},
	pages = {1266--1305},
}

@book{ruppert_semiparametric_2003,
	address = {Cambridge},
	series = {Cambridge {Series} in {Statistical} and {Probabilistic} {Mathematics}},
	title = {Semiparametric {Regression}},
	isbn = {978-0-521-78050-6},
	url = {https://www.cambridge.org/core/books/semiparametric-regression/02FC9A9435232CA67532B4D31874412C},
	urldate = {2021-02-12},
	publisher = {Cambridge University Press},
	author = {Ruppert, David and Wand, M. P. and Carroll, R. J.},
	year = {2003},
	doi = {10.1017/CBO9780511755453},
}

@article{gorecki_selected_2018,
	title = {Selected statistical methods of data analysis for multivariate functional data},
	volume = {59},
	issn = {1613-9798},
	url = {https://doi.org/10.1007/s00362-016-0757-8},
	doi = {10.1007/s00362-016-0757-8},
	language = {en},
	number = {1},
	urldate = {2021-02-10},
	journal = {Statistical Papers},
	author = {Górecki, Tomasz and Krzyśko, Mirosław and Waszak, Łukasz and Wołyński, Waldemar},
	month = mar,
	year = {2018},
	pages = {153--182},
}

@article{happ_multivariate_2018,
	title = {Multivariate {Functional} {Principal} {Component} {Analysis} for {Data} {Observed} on {Different} ({Dimensional}) {Domains}},
	volume = {113},
	issn = {0162-1459},
	url = {https://doi.org/10.1080/01621459.2016.1273115},
	doi = {10.1080/01621459.2016.1273115},
	number = {522},
	urldate = {2021-02-10},
	journal = {Journal of the American Statistical Association},
	author = {Happ, Clara and Greven, Sonja},
	month = apr,
	year = {2018},
	note = {Publisher: Taylor \& Francis
\_eprint: https://doi.org/10.1080/01621459.2016.1273115},
	pages = {649--659},
}

@misc{greven_denseflmm_2018,
	title = {{denseFLMM}: {Functional} {Linear} {Mixed} {Models} for {Densely} {Sampled} {Data}. {R} package version 0.1.2. {https://CRAN.R-project.org/pack} {age=denseFLMM}},
	copyright = {GPL-2},
	shorttitle = {{denseFLMM}},
	url = {https://CRAN.R-project.org/package=denseFLMM},
	urldate = {2021-02-09},
	author = {Greven, Sonja and Cederbaum, Jona},
	month = apr,
	year = {2018},
}

@book{faraway_extending_2016,
	title = {Extending the {Linear} {Model} with {R} : {Generalized} {Linear}, {Mixed} {Effects} and {Nonparametric} {Regression} {Models}, {Second} {Edition}},
	isbn = {978-1-315-38272-2},
	shorttitle = {Extending the {Linear} {Model} with {R}},
	url = {https://www.taylorfrancis.com/books/extending-linear-model-julian-faraway/10.1201/9781315382722},
	language = {en},
	urldate = {2020-12-21},
	publisher = {Chapman {\&} Hall/CRC},
	author = {Faraway, Julian J.},
	month = mar,
	year = {2016},
	doi = {10.1201/9781315382722},
}

@article{bates_fitting_2015,
	title = {Fitting {Linear} {Mixed}-{Effects} {Models} {Using} lme4},
	volume = {67},
	copyright = {Copyright (c) 2015 Douglas Bates, Martin Mächler, Ben Bolker, Steve Walker},
	issn = {1548-7660},
	url = {https://www.jstatsoft.org/index.php/jss/article/view/v067i01},
	doi = {10.18637/jss.v067.i01},
	language = {en},
	number = {1},
	urldate = {2020-12-21},
	journal = {Journal of Statistical Software},
	author = {Bates, Douglas and Mächler, Martin and Bolker, Ben and Walker, Steve},
	month = oct,
	year = {2015},
	note = {Number: 1},
	pages = {1--48},
}

@article{wu_predicting_2019,
	title = {Predicting fatigue using countermovement jump force-time signatures: {PCA} can distinguish neuromuscular versus metabolic fatigue},
	volume = {14},
	issn = {1932-6203},
	shorttitle = {Predicting fatigue using countermovement jump force-time signatures},
	url = {https://journals.plos.org/plosone/article?id=10.1371/journal.pone.0219295},
	doi = {10.1371/journal.pone.0219295},
	language = {en},
	number = {7},
	urldate = {2020-11-25},
	journal = {PLOS ONE},
	author = {Wu, Paul Pao-Yen and Sterkenburg, Nicholas and Everett, Kirsten and Chapman, Dale W. and White, Nicole and Mengersen, Kerrie},
	month = jul,
	year = {2019},
	note = {Publisher: Public Library of Science},
	pages = {e0219295},
}

@article{coffey_common_2011,
	title = {Common functional principal components analysis: {A} new approach to analyzing human movement data},
	volume = {30},
	issn = {0167-9457},
	shorttitle = {Common functional principal components analysis},
	url = {http://www.sciencedirect.com/science/article/pii/S0167945710001867},
	doi = {10.1016/j.humov.2010.11.005},
	language = {en},
	number = {6},
	urldate = {2020-09-23},
	journal = {Human Movement Science},
	author = {Coffey, N. and Harrison, A. J. and Donoghue, O. A. and Hayes, K.},
	month = dec,
	year = {2011},
	pages = {1144--1166},
}

@article{jacques_model-based_2014,
	title = {Model-based clustering for multivariate functional data},
	volume = {71},
	issn = {0167-9473},
	url = {http://www.sciencedirect.com/science/article/pii/S0167947312004380},
	doi = {10.1016/j.csda.2012.12.004},
	language = {en},
	urldate = {2020-09-22},
	journal = {Computational Statistics \& Data Analysis},
	author = {Jacques, Julien and Preda, Cristian},
	month = mar,
	year = {2014},
	pages = {92--106},
}

@article{liebl_ankle_2014,
	title = {Ankle plantarflexion strength in rearfoot and forefoot runners: {A} novel clusteranalytic approach},
	volume = {35},
	issn = {0167-9457},
	shorttitle = {Ankle plantarflexion strength in rearfoot and forefoot runners},
	url = {http://www.sciencedirect.com/science/article/pii/S0167945714000402},
	doi = {10.1016/j.humov.2014.03.008},
	language = {en},
	urldate = {2020-09-16},
	journal = {Human Movement Science},
	author = {Liebl, Dominik and Willwacher, Steffen and Hamill, Joseph and Brüggemann, Gert-Peter},
	month = jun,
	year = {2014},
	pages = {104--120},
}

@article{ryan_functional_2006,
	title = {Functional data analysis of knee joint kinematics in the vertical jump},
	volume = {5},
	issn = {1476-3141},
	doi = {10.1080/14763141.2006.9628228},
	language = {eng},
	number = {1},
	journal = {Sports Biomechanics},
	author = {Ryan, Willie and Harrison, Andrew and Hayes, Kevin},
	month = jan,
	year = {2006},
	pmid = {16521626},
	pages = {121--138},
}








\end{document}